\documentclass[
 reprint,
superscriptaddress,
nofootinbib,
 amsmath,amssymb,
 aps,prc
]{revtex4-2}

\usepackage{graphicx}
\usepackage{dcolumn}
\usepackage{bm}

\usepackage[draft, inline, nomargin]{}

\newcommand{\Bten}{$^{10}$B}
\newcommand{\Lisix}{$^{6}$LiF}
\newcommand{\Hefour}{$^{4}$He}
\newcommand{\Htwo}{H$_{2}$}
\newcommand{\Htwop}{H$_{2}^+$}
\newcommand{\HtwoO}{H$_{2}$O}
\newcommand{\COtwo}{CO$_{2}$}
\newcommand{\microgcm}{$\mu$g/cm$^2$}

\begin{document}

\preprint{APS/123-QED}

\title{Detection of Molecular Hydrogen in a Neutron Beam Lifetime Experiment}

\author{J. Caylor}
\affiliation{Thomas Jefferson National Accelerator Facility, Newport News, VA 23606 USA}
\affiliation{University of Tennessee, Knoxville, TN 37996 USA}
\author{R. Biswas}
\affiliation{Tulane University, New Orleans, LA, 70118, USA}
\author{B. Crawford}
\affiliation{Gettysburg College, Gettysburg, PA, 17325, USA}
\author{M.S. Dewey}
\affiliation{National Institute of Standards and Technology, Gaithersburg, MD 20899 USA}
\author{N. Fomin}
\affiliation{University of Tennessee, Knoxville, TN 37996 USA}
\author{G.L. Greene}
\affiliation{University of Tennessee, Knoxville, TN 37996 USA}
\affiliation{Tulane University, New Orleans, LA, 70118, USA}
\author{S.F. Hoogerheide}
\email{Contact author: shannon.hoogerheide@nist.gov}
\affiliation{National Institute of Standards and Technology, Gaithersburg, MD 20899 USA}
\author{J. Hungria-Negron}
\affiliation{Gettysburg College, Gettysburg, PA, 17325, USA}
\author{H.P. Mumm}
\affiliation{National Institute of Standards and Technology, Gaithersburg, MD 20899 USA}
\author{J.S. Nico}
\affiliation{National Institute of Standards and Technology, Gaithersburg, MD 20899 USA}
\author{F.E. Wietfeldt}
\affiliation{Tulane University, New Orleans, LA, 70118, USA}
\author{D.O. Valete}
\affiliation{Gettysburg College, Gettysburg, PA, 17325, USA}
\author{J. Zuchegno}
\affiliation{Tulane University, New Orleans, LA, 70118, USA}

\date{\today}

\begin{abstract}
One method of determining the free neutron lifetime involves the absolute counting of neutrons and trapped decay protons. In such experiments, a cold neutron beam traverses a segmented proton trap inside a superconducting solenoid while the neutron flux is continuously monitored. Protons that are born within the fiducial volume of the trap are confined radially by the magnetic field and axially by the electrostatic potential supplied by trap electrodes. They are periodically released and counted, and the ratio of the absolute number of neutrons to protons is proportional to the neutron lifetime. Systematic error can be introduced if protons in the trap are lost, gained, or misidentified. The influence of molecular hydrogen interactions is of particular interest because of its ubiquitous presence in ultrahigh vacuum systems.  To understand how it could affect the neutron lifetime, measurements were performed on the production and detection of molecular hydrogen in an apparatus used to measure the neutron lifetime. We demonstrate that charge exchange with molecular hydrogen can occur with trapped protons, and we determine the efficiency with which the molecular hydrogen ions in the trap are detected. Finally, we comment on the potential impact on a neutron lifetime experiment using this beam technique. We find that the result of the beam neutron lifetime performed at NIST is unlikely to have been significantly affected by charge exchange with molecular hydrogen.
\end{abstract}

\maketitle


\section{\label{sec:overview}Neutron lifetime measurement}

The free neutron is unstable and decays into a proton, an electron, and an antineutrino. Several experimental techniques have been developed over many decades to measure its lifetime. One such technique, the ``beam method," counts the decay products from a beam of slow neutrons while also measuring the number of neutrons in the decay volume. Another commonly used method, the ``bottle method", confines ultracold neutrons (UCNs) in a material or magnetic bottle and counts the number remaining after varying storage times.   The motivation for measuring the neutron lifetime along with details on the history of lifetime measurements and the different measurement techniques can be found in several reviews~\cite{Wietfeldt_2011, Dubbers_2011, Dubbers_2021}.

The most precise determinations of the neutron lifetime have come from recent UCN magnetic bottle experiments, which have achieved 0.3\,s uncertainty~\cite{Pattie_2018,Gonzalez_2021,Musedinovic_2025} while the most precise beam measurement has an uncertainty of 2.2\,s~\cite{Yue2013}. The agreement between the beam measurements and the recent UCN experiments is poor with a difference of over 4\,$\sigma$~\cite{Wietfeldt_2011,Workman_2022}. To address this difference, efforts are underway to perform new measurements of the neutron lifetime using the beam method~\cite{Hoogerheide2019,Hirota2020,Fuwa2024,BL3_2021}, improving the precision using UCN~\cite{FunPhysWhite2023}, as well as pursuing novel techniques~\cite{Materne2009,Hassan2021,Lawrence2021}. Much work has also gone into exotic explanations~\cite{Fornal2018,Cline2018,Rajendran2021,Koch2024,Kading_2025}, but as yet, none has compelling support~\cite{Baym2018,Motta2018,McKeen2018,Tang2018,Sun2018,Dubbers2019,Klopf2019,Broussard2022,Joubioux2024,Blatnik2024}. Because the beam measurement average is significantly discrepant in comparison with the more recent UCN measurements, it is important to examine systematic effects that could affect a beam measurement.

A beam measurement of the neutron lifetime requires the absolute counting of neutrons and at least one of the decay products. While many of the pioneering experiments have used that general concept, the most precise measurements have utilized a quasi-Penning trap to confine decay protons and a thin \Bten\ or \Lisix\ deposit viewed by particle detectors to measure the neutron beam density. The first such experiment was performed at the Institut Laue–Langevin (ILL)~\cite{Byrne_1990,Byrne_1996} and a similar experiment was later carried out at the National Institute of Standards and Technology (NIST)~\cite{Dewey_2003,Nico_2005}. There are challenges measuring the lifetime with the beam method. The relatively long neutron lifetime means there are few decay protons to detect; a precise measurement of the neutron decay volume must be made; and the density of the neutrons in the trapping region must be measured. As such, good knowledge of the absolute efficiencies of both proton and neutron detection is essential.

Given the situation with recent lifetime measurements, carefully revisiting all systematic effects is well motivated. Of particular interest for beam measurements is understanding systematic effects that would increase the lifetime. For experiments that count protons, any systematic effect that could lead to an unaccounted loss of protons would give a value of the neutron lifetime that is erroneously high. One potential effect for experiments that trap the decay protons is the interaction of residual gas with the  protons~\cite{Byrne_2019,Byrne_2022,Serebrov_2021,Wietfeldt_2022}. In this paper, we focus on the results of studies of proton loss mechanisms related to hydrogen gas interactions and their possible effects on the measured value of the neutron lifetime.

\section{\label{sec:ResidualGas}Residual Gas Interactions in the Proton Trap}

The apparatus for this work was mounted at the NIST Center for Neutron Research on the NG-C beamline in the cold neutron guide hall. Its main components are a cold-neutron beamline, a neutron flux monitor, a segmented proton trap inside the bore of a superconducting magnet, and the associated counting electronics. The operation of the trap and detector is discussed in more detail in Section~\ref{sec:expt}.

\begin{figure}[ht]
    \centering
    \includegraphics[width=.45\textwidth]{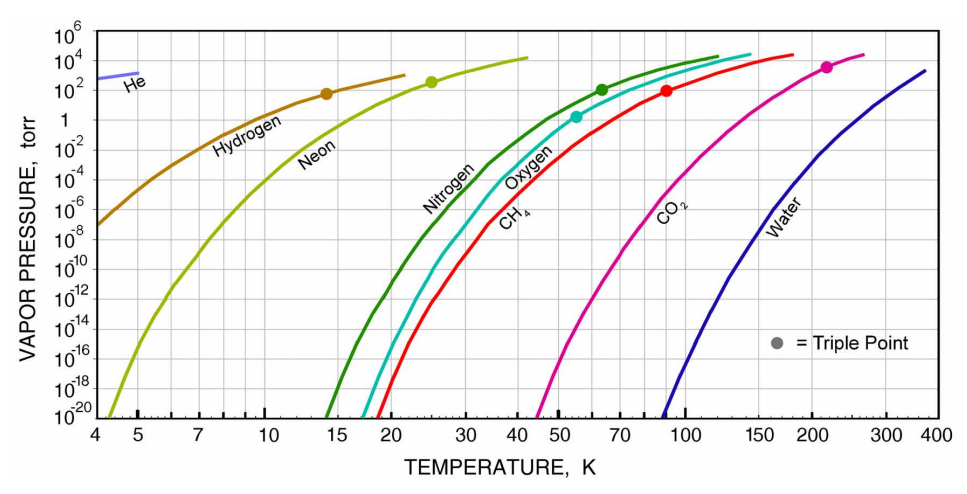}
    \caption{Saturated vapor pressure of common gases as a function of temperature. (Reprinted with permission from Ref.~\cite{BarCohen2016}.)}
    \label{fig:vapor_pressure}
\end{figure}

Despite the lack of direct measurements of residual gas pressures in the region of the proton trap, it is possible to constrain partial pressures based on features of the apparatus. The proton trap resided in the bore of a superconducting magnet that was cooled by liquid helium.  The operational temperature of the bore was measured to be less than 10\,K although the trap itself was approximately 40\,K due to the weak thermal contact between the trap and bore. As discussed in Ref.~\cite{Wietfeldt_2022}, the magnet bore was effectively a cryopump and, as such, the partial pressure of all residual gases other than helium, hydrogen, and neon is considered to be negligible. This can be seen from Fig.~\ref{fig:vapor_pressure}, which shows the calculated saturated vapor pressure for many common gases as a function of temperature~\cite{BarCohen2016}. There is no reason to expect any neon inside the vacuum system although hydrogen is present and possibly helium due to its use in the magnet and guide hall.

Protons confined in the trap are subject to three types of interactions with residual gas in the trapping volume: elastic scattering, inelastic scattering, and charge exchange. The first two types of scattering do not change the efficiency of proton detection because they do not increase the proton energy as the residual gas molecules have energies on the order of meV. As such, they are not expected to have a significant effect on the measured neutron lifetime. The charge exchange interaction dominates in the low energy regime where the proton velocity is less than the orbital electron velocity of the residual gas atom (molecule) \cite{Byrne_2019}. The charge exchange reaction of a trapped proton and a residual gas atom (molecule) is given by

\begin{equation}\label{eqn:chexchange}
	{\rm {p + M \rightarrow H + M^+}},
\end{equation}

\noindent where p is a proton, M is any atom or molecule, and H is a neutral hydrogen atom. The proton captures an electron from a residual gas atom (molecule) and is converted to a neutral hydrogen atom, which escapes the trap, and the new singly charged atom (molecule) is left trapped. The M$^+$ ion then acts in a manner similar to a trapped proton, confined axially by the electrostatic field of the trap electrodes and radially by the magnetic field. When the trap transitions into counting mode, the M$^+$ ion is transported to the proton detector by the acceleration potential (as described in Section~\ref{subsec:ProtonTrap}) and is detected with an efficiency that, in principle, varies with each ion species.

The probability, \(P_{M}(x)\), of proton charge exchange occurring is

\begin{equation}\label{eqn:chexchangeXsection}
P_{M}(x) = 1 - e^{-n\sigma_Mx} \approx n\sigma_M x,
\end{equation} 

\noindent where $n$ is the number density of the residual gas during a trapping cycle, \(\sigma_M\) is the charge exchange cross section for each residual gas species, and $x$ is the distance traveled by the proton through the residual gas~\cite{Byrne_2019}. At low energies, $< 1$\,keV, charge exchange cross sections vary significantly over the range of light elements and common molecules \cite{AtomicData_1978,Phelps_1992,Berkner_1970,Stedeford_1955,Gilbody_1957,Stebbings_1964,Allison_1958}. The charge exchange cross section for \Hefour\ is approximately two orders of magnitude lower than molecular hydrogen for the relevant energy ranges \cite{Phelps_1992,Allison_1958,Stedeford_1955}, and we assume the partial pressure of all other gases are extremely low. Consequently, the focus of this paper will be on the charge exchange due to molecular hydrogen, \Htwo. We note that it is possible to produce {H$_{3}^+$} ions via secondary reactions of \Htwop\ with \Htwo~\cite{Byrne_2022}, but it is with a lower probability. It is also possible that $\text{He}^+$ ions would also be detectable in our apparatus, but based on simulation, they would appear at a slightly higher energy relative to protons due to losing less energy in the detector deadlayer. No such signal has been observed in data acquired for these studies or the 2003 experiment~\cite{Dewey_2003}.

\section{\label{sec:expt}Experimental Approach}
\subsection{\label{subsec:apparatus}Overview of the Apparatus}

The apparatus for these studies is part of the BL2 experiment to measure the neutron lifetime~\cite{Hoogerheide2019}. It is largely the same as that used to perform the lifetime measurement at NIST in 2003. The apparatus is very similar in concept and design to that used in the measurement at the ILL in 1989~\cite{Williams1989,Byrne1989}.  We review some of the aspects that are relevant to this work, and additional details of its operation are found in Refs.~\cite{Nico_2005,Caylor2022}.

\begin{figure}[ht]
    \centering
    \includegraphics[width=.45\textwidth]{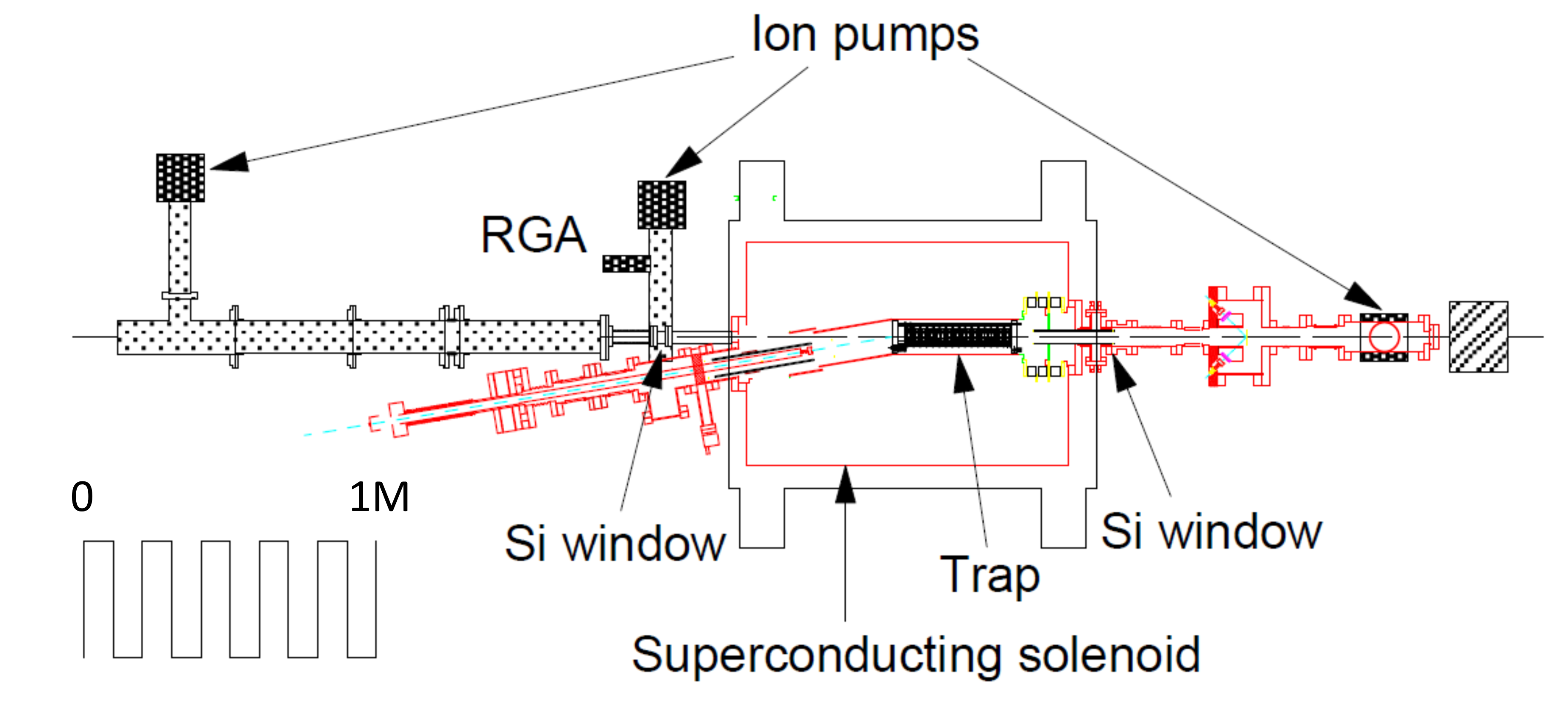}
    \caption{Layout of the apparatus for this work. The neutron beam travels left to right and is detected downstream from the trap.  Decay protons are detected upstream of the trap near the bend.  Although it is one contiguous vacuum system, there are two thin, silicon windows that can isolate the bore from the rest of the vacuum system.}
    \label{fig:BL2beamline}
\end{figure}

Figure~\ref{fig:BL2beamline} shows the layout of the cold neutron beamline and the counting apparatus. The proton trap lies inside the bore of the superconducting solenoid. A silicon detector (300\,mm$^2$ or 600\,mm$^2$) is mounted on the end of a tube that is $9.5^{\circ}$ from the bore axis. It can be moved in and out of the solenoid via a translation stage. During operation, the detector is moved into its measurement position in the bore and cools to a temperature of about 150\,K, reducing the leakage current and noise of the detector. It can be pulled back to a position at room temperature where it can be serviced. The pressure is maintained by three ion pumps placed on the beamline, near the magnet bore, and at the neutron flux monitor. 

Importantly, two thin, perfect-crystal silicon windows are located just upstream and downstream of the magnet bore. These windows are mounted on gate valves so that they can be inserted and removed \textit{in situ} to systematically vary the vacuum conditions inside the magnet bore and study the effects of gaseous species in the trap. The thin windows allow the cold bore of the magnet to be better isolated from the room-temperature flight tubes upstream and downstream of the magnet. This results in reduced pressure in those regions from which one infers better vacuum conditions in the trapping region. It also increases the time the trap can be stably held in the trapping configuration by over an order of magnitude~\cite{Caylor2022}.

The number of neutrons is determined using a neutron flux monitor downstream of the magnet. The monitor consists of an array of four silicon detectors surrounding a thin \Lisix\ deposit. The \Lisix\ has a neutron capture cross-section that is inversely proportional to the neutron's velocity over the relevant cold neutron energies. When a neutron is captured by the \Lisix, the $^{7}$Li promptly decays into an alpha particle and a triton. These particles have energies of a few MeV and are easily detectable with silicon detectors. Only a small fraction of the total number of neutrons are detected this way, so for a lifetime measurement, the detection efficiency must be determined to find the absolute number of neutrons in the beam~\cite{Nico_2005,Yue_2018}.

The absolute number of neutrons is not critical for these studies of proton loss mechanisms, but a neutron flux measurement is essential to normalize the measured proton rates. For completeness, the lifetime $\tau_n$ is extracted by determining \(\Dot{N}_p/\Dot{N}_{\alpha +t}\) as a function of the trap length, where $\Dot{N}_p$ is the measured proton rate and $\Dot{N}_{\alpha +t}$ is the measured rate of alpha and triton particles from the neutron flux monitor, a rate that is proportional to the neutron flux.  The neutron lifetime is given by

\begin{equation}\label{eq:lifetime}\frac{\dot{N}_p}{\dot{N}_{\alpha+t}} = \tau^{-1}_n \left(\frac{\epsilon_p}{\epsilon_o v_o}\right)(nl + L_{end}),
\end{equation}

\noindent where \(\epsilon_p\) and \(\epsilon_o\) are the proton and neutron detection efficiencies, \(v_0\) is the defined thermal neutron velocity (2200\,m/s), \(n\) is the number of trapping electrodes, \(l\) is the length of the electrode and its adjacent spacer, and \(L_{end}\) is the effective length of the end regions of the proton trap~\cite{Nico_2005}. The decay probability inside the trapping region and the neutron capture probability by the thin \Lisix\ foil are both inversely proportional to the neutron's velocity, so precise knowledge of the neutron wavelength spectrum is not needed.

\subsection{\label{subsec:ProtonTrap}Proton Trap}

\begin{figure}[ht]
    \centering
    \includegraphics[width=.45\textwidth]{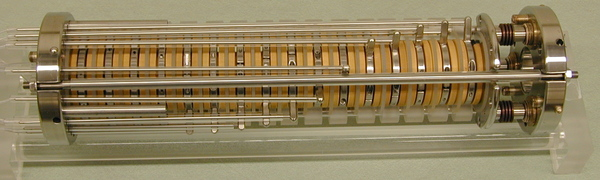}
    \caption{Photo of the segmented electrode proton trap.}
    \label{fig:trap}
\end{figure}

The proton trap consists of 16 cylindrical electrodes and was used in both this work and the previous lifetime measurement at NIST~\cite{Dewey_2003,Nico_2005}. The electrodes are made from fused quartz and coated with a thin layer of gold to make them electrically conductive. The electrodes are mounted with (uncoated) fused quartz spacers between each electrode to make a segmented proton trap, as shown in Fig.~\ref{fig:trap}. Both the dimensions of the trap and its position within the magnet bore are precisely known.

\begin{figure}[ht]
    \centering
    \includegraphics[width=.45\textwidth]{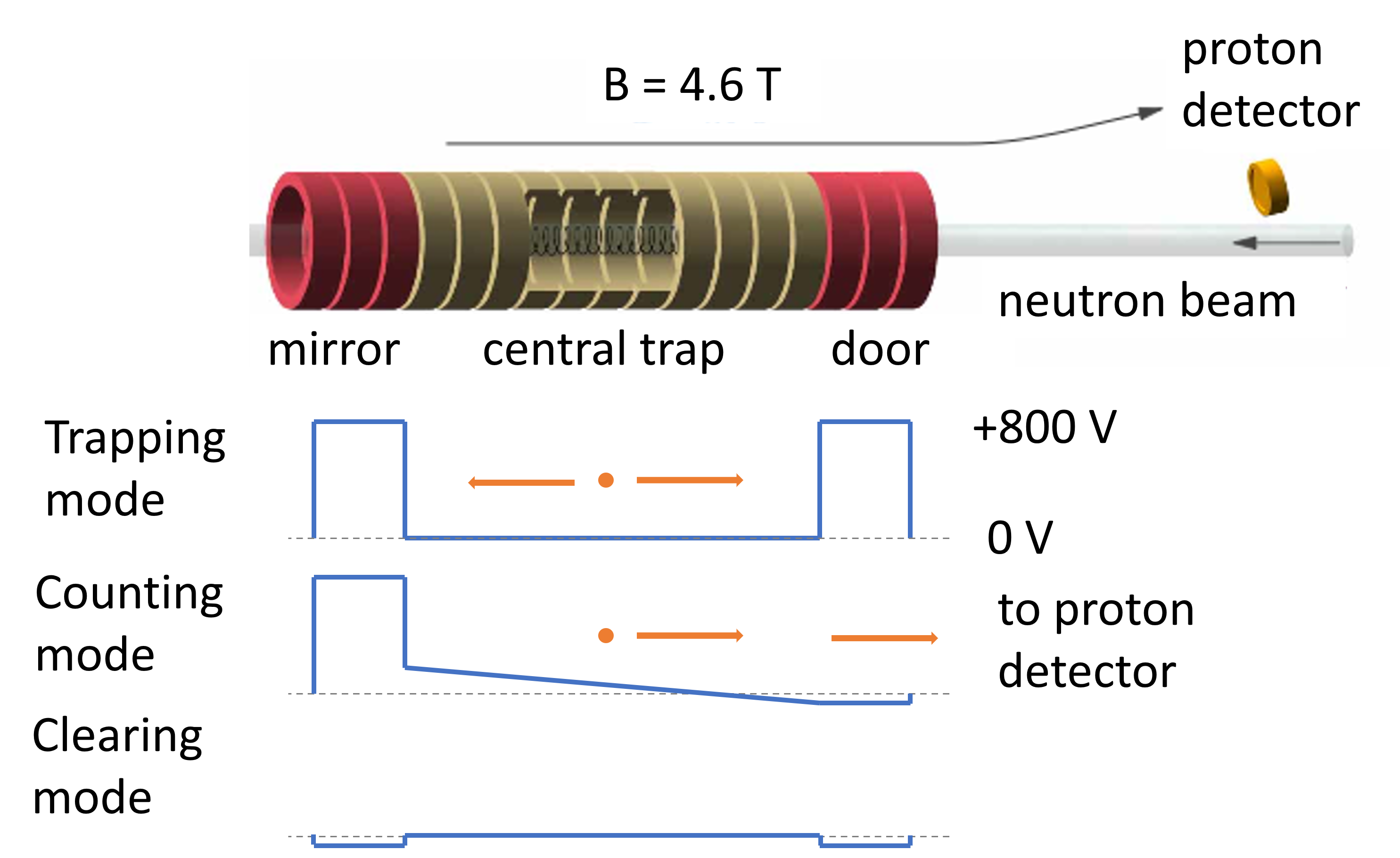}
    \caption{Schematic of the proton trap and detector (top) during a trapping mode. The diagrams below it illustrate the electrostatic potentials in the trap during the three modes of a trapping cycle.}
    \label{fig:trapcycle}
\end{figure}

Electrostatic potentials are applied to the electrodes to perform different functions: trapping, counting, and clearing, as shown in Fig.~\ref{fig:trapcycle}. In trapping mode, the first three electrodes (upstream relative to the beam travel) are energized and form the ``door'' of the trap; they are fixed in position and do not move. The central region of the trap is variable in length and is kept grounded. The ``mirror'' is formed by three energized electrodes and immediately follows the last (furthest downstream) electrode of the central trapping region; its position varies depending on the length of the trapping region. Both the door and mirror electrodes are held at a voltage (typically $+800$\,V) large enough to fully trap the most energetic decay protons.  A proton that decays in the central region is trapped axially by the electrostatic potential and radially by a 4.6\,T magnetic field.

Because of the shape of the electrostatic potential, there exist regions close to the door and mirror where protons are not trapped with 100\,\% efficiency, these are called the end regions ($L_{end}$ of Eq.~\ref{eq:lifetime}). For this reason, the length of the central trapping region is varied during data taking. The precisely manufactured electrodes keep $L_{end}$ constant for all trapping configurations. Varying the central trapping region allows the effect of $L_{end}$ to be extrapolated out in the neutron lifetime, as seen in Eq.~\ref{eq:lifetime}. Any electrodes that are downstream of the mirror are held at ground potential.

In counting mode, the voltage on the door is lowered and a small gradient ``ramp'' potential is applied to the central region to eject the protons from the trap. The protons follow the magnetic field lines and are accelerated by the high negative voltage applied to the proton detector. This high voltage potential is needed because the low-energy decay protons do not have enough energy to penetrate the deadlayer of typical silicon detectors. For data in this work, the acceleration potential is $-25$\,kV unless noted otherwise.

The clearing mode is the last configuration of electrode potentials. Potentials on the door and mirror are set to a slightly negative voltage to ensure that negatively charged particles are flushed from the entire trapping region. Upon completion of the clearing mode, another trapping cycle begins.

\subsection{\label{subsec:DAQ}Data Acquisition and Analysis}
\subsubsection{\label{subsubsec:ProtonSpectra}Proton Energy and Timing Spectra}

The proton electronics chain starts with the silicon detector. There are two commonly used silicon detector types for proton detection: passivated implanted planar silicon (PIPS), which has some amount of silicon dioxide on the entrance window, and surface barrier (SB), which is available with varying thicknesses of gold as the entrance window. These entrance windows are referred to as deadlayers because they do not contribute to the measured signal. They each have different energy loss characteristics and backscattering probabilities that must be well understood for measuring the neutron lifetime, as well as for other neutron decay experiments.

The silicon detector signal was read by a low-noise preamplifier made specifically for this experiment. The output of the preamplifier was sent via a fiber optic cable into a NIM module and split into two parallel signals. One signal was sent directly into one channel of a 12-bit, 500 megasample per second digitizer. The waveform was digitized starting approximately $50\,\mu$s before the door opened and ending approximately $150\,\mu$s after the door opened. Depending on the ramp voltage, the protons arrived up to about $20\,\mu$s after the door opened. The waveform was analyzed using two trapezoidal filters, one optimized for energy resolution and one for timing resolution~\cite{Caylor2022,trapfilter}. Using the optimized trapezoid filters, the energy and arrival time of events relative to the trap opening were obtained.

\begin{figure}[ht]
    \centering
    \includegraphics[width=8.6 cm]
    {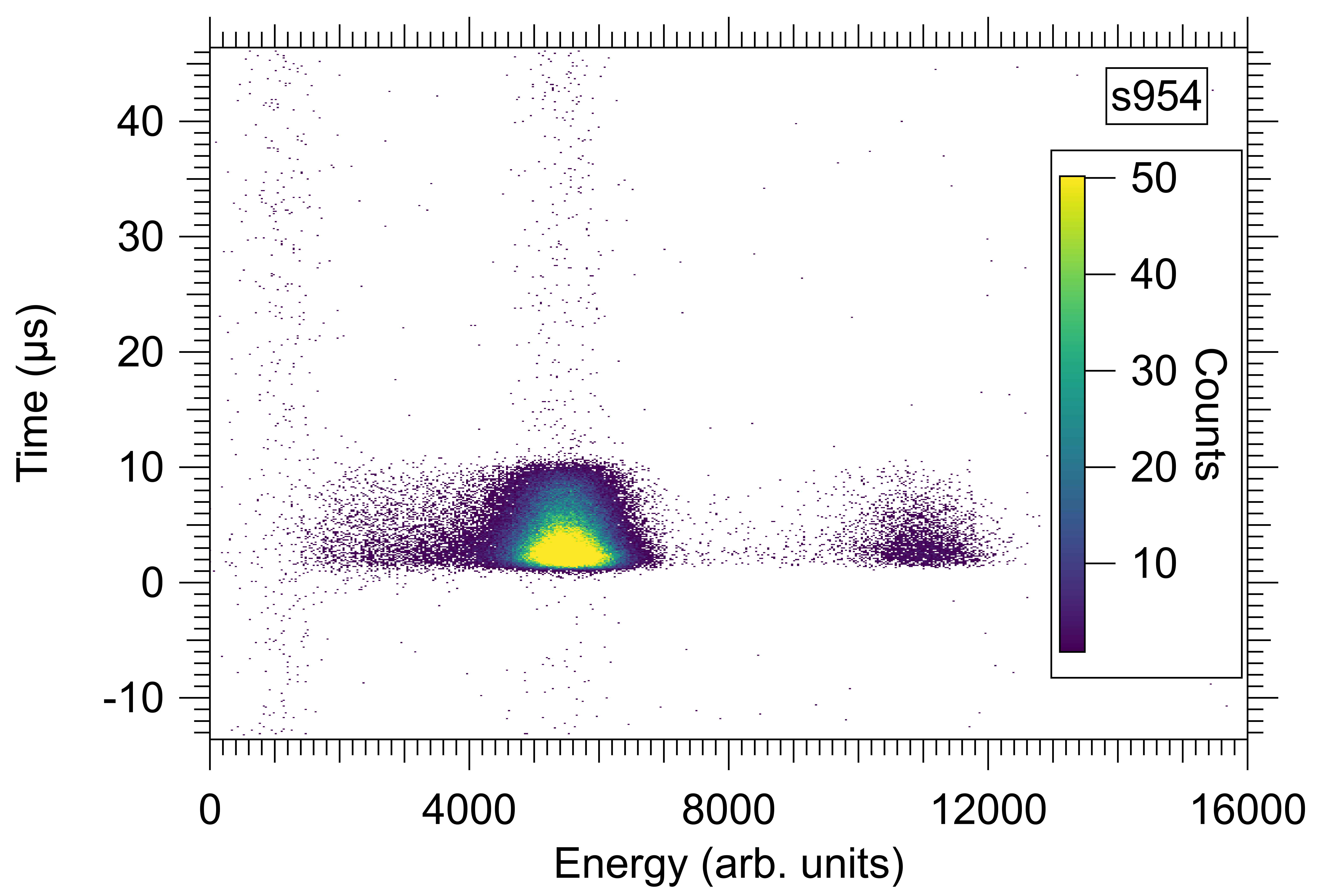}
    \caption{2-dimensional histogram of the arrival time versus the energy after the door voltage of the trap is lowered.}
    \label{fig:2DPlot}
\end{figure}

Figure~\ref{fig:2DPlot} shows an example of a 2-dimensional histogram of event arrival time and energy. The protons arrived at the detector within approximately $10\,\mu$s after the trap was opened. The main features are the intense peak of single proton events, the less intense peak at twice the energy where two protons were trapped, and the band of so-called inflight protons resulting from neutrons that decay randomly along the trap while the door is open. The peak with two protons is small relative to the main peak due to the low counting rate (a few per second), but multiple proton events must be accounted for in a lifetime measurement. The low-energy tail on the protons is also an important feature; it results predominantly from the energy loss of some protons as they scatter within the detector deadlayer. When measuring the lifetime, all the protons must be accounted for, including those that are in the tail, fall below the energy threshold, stop in the deadlayer, or backscatter.

\section{\label{sec:H2}Properties of Trapped \Htwop\ Ions} 

Proton charge exchange with molecular hydrogen (\Htwo) producing singly ionized molecular hydrogen (\Htwop) could result in observable effects under the appropriate conditions. In this section, we present methods by which \Htwop\ ions created via charge exchange can be identified in our data. Evidence includes timing signatures of the low-energy trapped ions (Section~\ref{subsec:LowEnergyIons}), pressure and trap-time dependence (Section~\ref{subsec:Pressure}), and comparison of data with timing simulations of \Htwop\ trapping and transport (Section~\ref{subsec:ArrivalTime}).  The data for this work were acquired between July 2018 and October 2020. The data in this period were part of studies re-examining systematic effects associated with beam experiments using this technique.

\subsection{\label{subsec:LowEnergyIons}Timing of \Htwop\ Ions}

A trapped proton undergoing the charge-exchange reaction \Htwo[p,\Htwop]H would result in the \Htwop\ ion being trapped while the H atom escapes from the trap. With a sufficient density of \Htwo\ molecules, it should be possible to detect the \Htwop\ ion from this reaction with this apparatus using its energy and timing signatures. An \Htwop\ ion at room temperature has an energy that is a small fraction of an eV. If trapped, these ions have much lower energies than decay protons, which have energies up to the endpoint of 782\,eV. Consequently, they will have different timing and energy structures after the trap is opened and the ejected particles follow the magnetic field lines to the detector.

\begin{figure}[h!]
    \centering
    \includegraphics[width=8.6 cm] {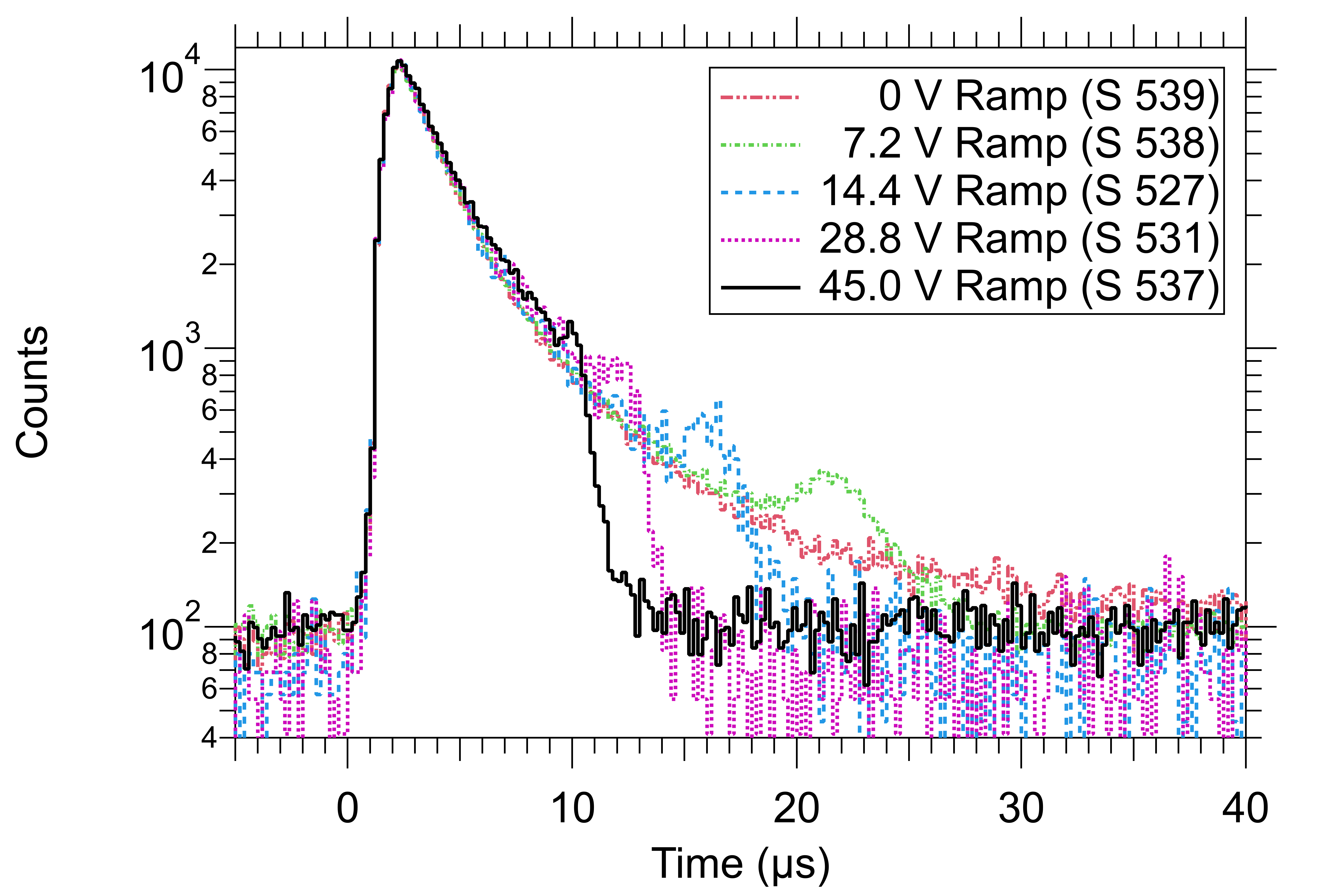}
    \caption{Experimental data showing the arrival times of the proton and \Htwop\ peaks with varied ramp voltage. Lower ramp voltages lead to a later arrival time and easier identification of the \Htwop\ peak.  All data were obtained using a trap length of 9 electrodes; the stated ramp voltage corresponds to the voltage on the 9th electrode.}
    \label{fig:arrival_bonus_peak_ramp_voltage}
\end{figure}

The majority of the decay protons have enough energy to leave the trap without the need for the gradient potential of the ramp. Those protons arrive at the detector a few microseconds after the door is opened, and the process continues until the low-energy protons are ejected by the ramp. Although the width of the proton time-of-flight peak is affected by the ramp, the initial arrival time is not. Because the \Htwop\ ions are very low energy, one does not expect to see these types of events arriving with the fastest protons. The vast majority of the ions are accelerated by the ramp, and the arrival time of these events depends almost entirely on the ramp voltage. This behavior is seen in the late-time peaks in the timing spectra of Fig.~\ref{fig:arrival_bonus_peak_ramp_voltage}. The acquired data had a sufficient density of \Htwo\ molecules to create a measurable number of \Htwop\ ions. The plot shows the arrival time spectra for five different ramp voltages. As expected, the early-time timing structure is similar for all the ramp voltages. The highest energy decay protons arrive quickly and form the majority of the timing peak. The bump at the end of the timing spectra is what we attribute to the lower energy \Htwop\ ions. As the ramp voltage decreases, the \Htwop\ peak moves later in time and becomes more visible because nearly all decay protons arrive before the \Htwop. At zero ramp voltage, there is no \Htwop\ peak because there is no longer any potential to bunch the ions in time, and both the lowest-energy decay protons and the ions drift to the detector over longer times.

\begin{figure}[ht]
    \centering
    \includegraphics[width=.45\textwidth]{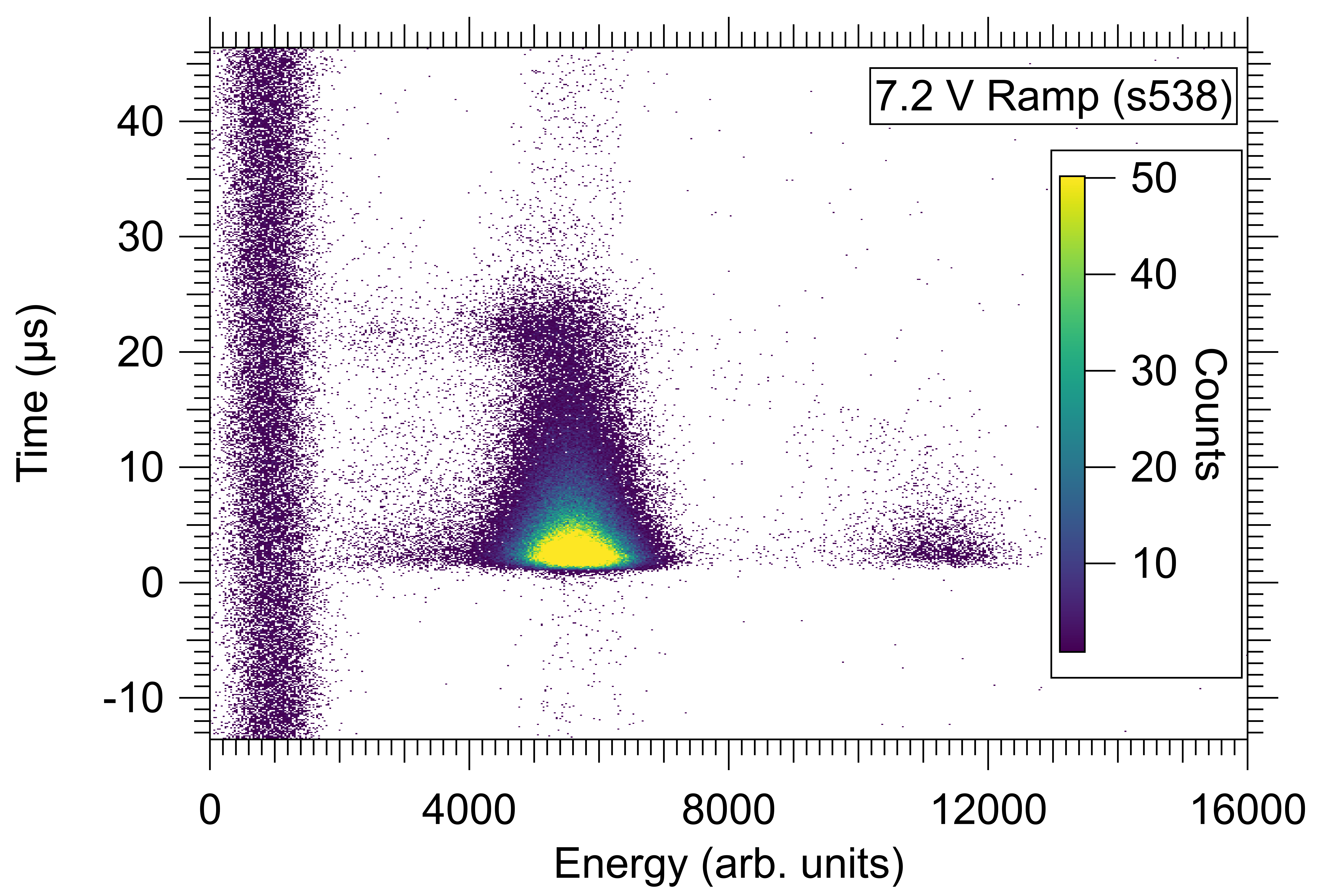}
    \includegraphics[width=.45\textwidth]{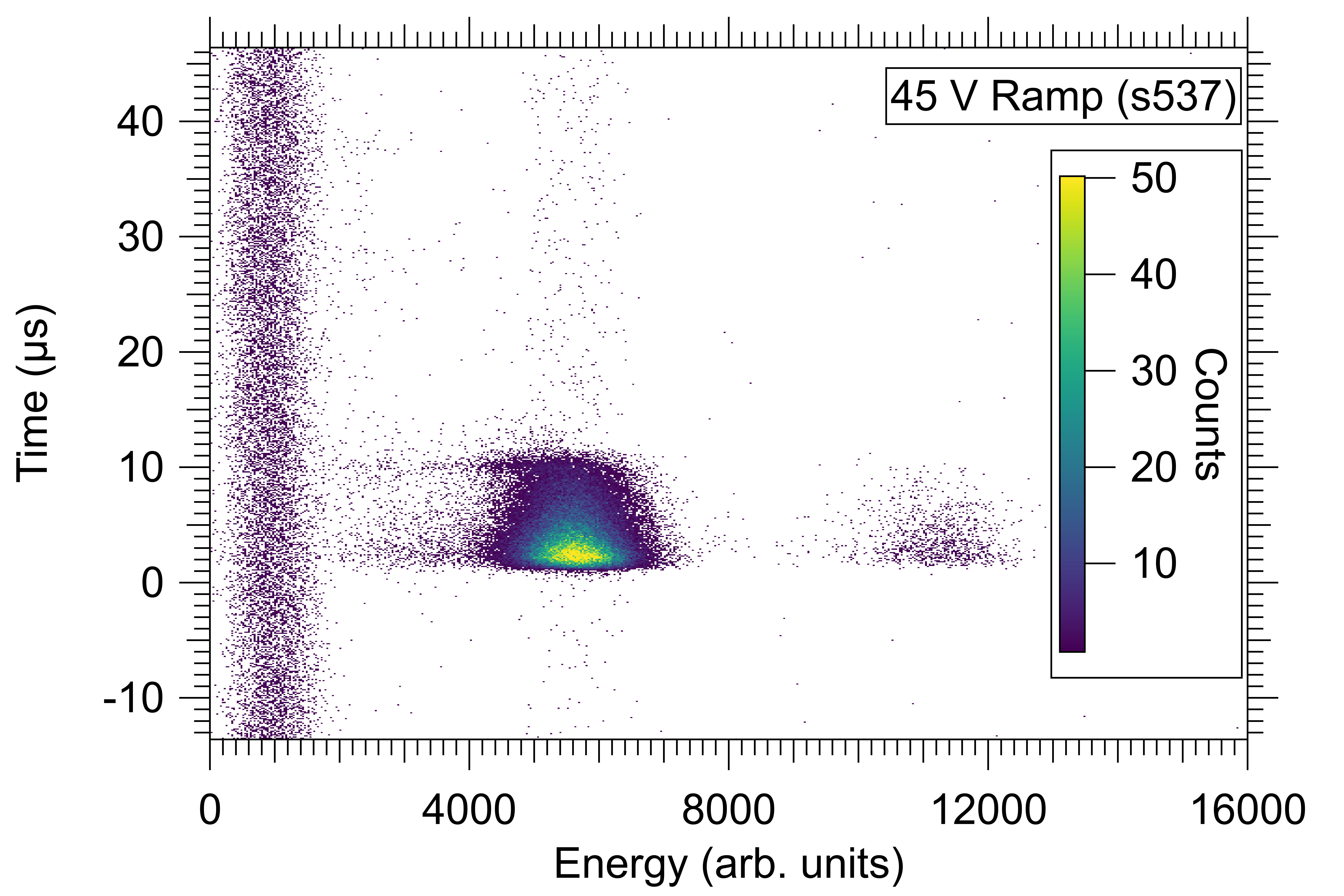}
    \caption{2-dimensional histograms of the arrival time versus the energy after the door voltage of the trap is lowered for two different ramp voltages. The top plot has a 7.2\,V ramp, and the bottom plot has a 45\,V ramp. The plots are similar to Fig.~\ref{fig:2DPlot} with the exception of the events occurring at late arrival times and at lower energies. These are events consistent with being \Htwop\ ions. They occur at later times than the main proton peak and have a different energy structure. The two plots illustrate the difference in the arrival time of the \Htwop\ ions depending on the ramp voltage. }
    \label{fig:2D_ramp_voltage_comp}
\end{figure}

Figure~\ref{fig:2D_ramp_voltage_comp} shows 2-dimensional histograms of proton energy and arrival time for two different ramp voltages (7.2\,V and 45\,V). The \Htwop\ events that are found peaked at late arrival times in Fig.~\ref{fig:arrival_bonus_peak_ramp_voltage} are distinct in the 2-dimensional histograms. One sees an excess number of events at the tail end of the trapped proton peak, and the position in time moves relative to the proton peak when the ramp voltage is changed. This behavior is consistent with the arrival times of \Htwop\ as seen in simulation. 

\subsection{\label{subsec:Pressure}Pressure and Trap Time Dependence}

The number of charge exchange events that one may detect is dependent on the density of the residual gas. As the partial pressure of a gas increases, the probability of charge exchange occurring also increases. To obtain an estimate for the proton loss probability from \Htwo, one must integrate Eq.~\ref{eqn:chexchangeXsection} over the charge-exchange cross section~\cite{Phelps_1990}, the proton velocity, and the energy spectrum of the decay protons. For a typical trap time of 10\,ms and assuming a trap temperature of 40\,K and a pressure of $1\times 10^{-7}$\,Pa, one obtains a loss probability of about 0.3\,\%. This value is consistent with an estimate from Ref.~\cite{Byrne_2019}. There is significant uncertainty in the estimate because residual \Htwo\ gas is unlikely to be in thermal equilibrium with the trap. In addition, the partial pressure of \Htwo\ is difficult to estimate precisely because the trap is weakly coupled to the upstream and downstream vacuum systems. 

\begin{figure}[h!]
    \centering
    \includegraphics[width=.45\textwidth]{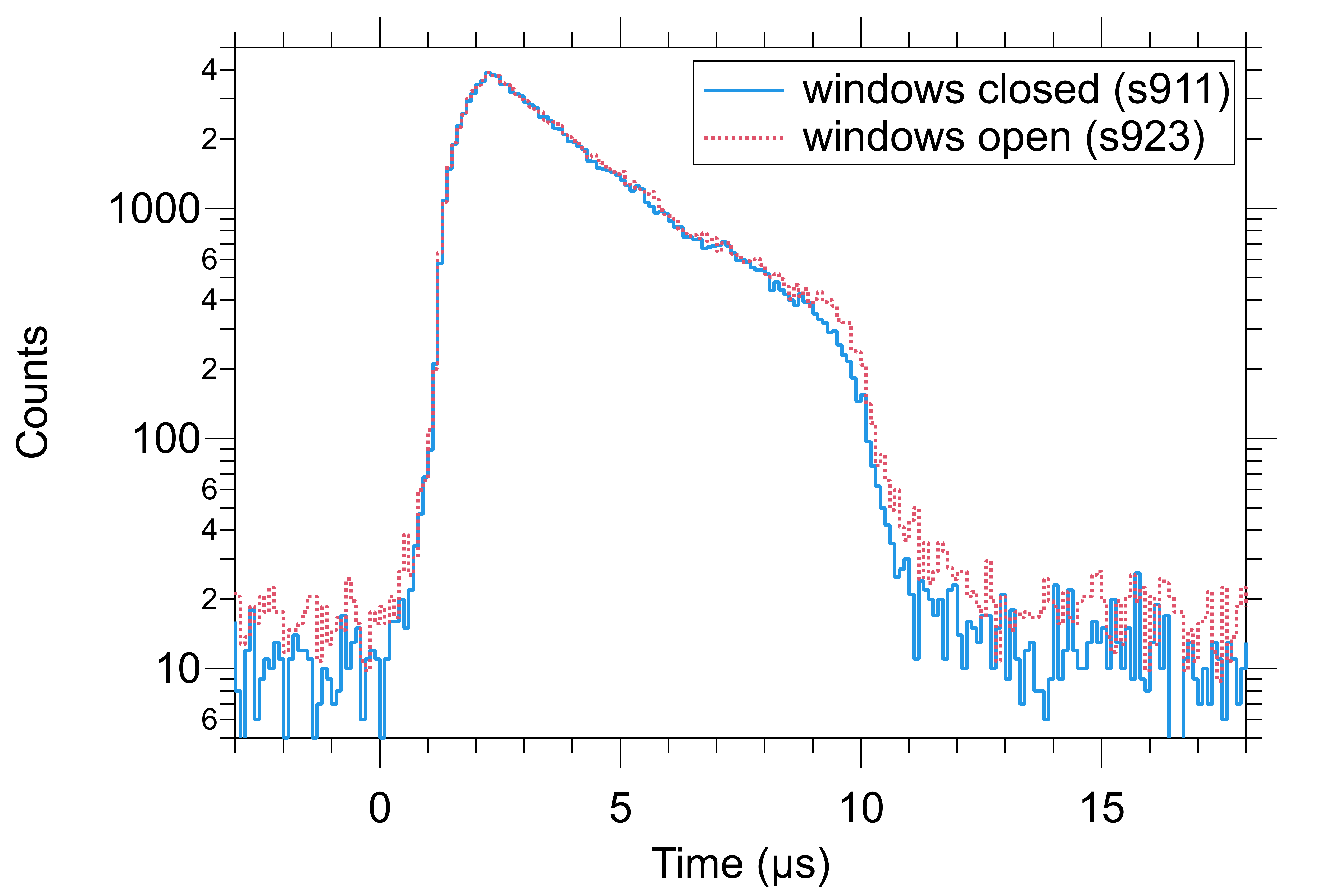}
    \caption{Comparison of proton arrival time spectra between Si windows open (red) and Si windows closed (blue). There is an increase in the number of events in the peak attributable to \Htwop\ at \(\approx 9.5\ \mathrm{\mu s}\) when the Si windows are open.}
    \label{fig:windows_open_closed_timing_spectrum}
\end{figure}

The silicon windows in the apparatus allowed isolation of the cold bore from the sections of the beamline that were not cryopumped, thus changing the density of \Htwo\ at the trap. When the silicon windows are closed, isolating the proton region from the beamline, the pressure improves by about a factor of three at the pressure gauge closest to the proton detector. Figure~\ref{fig:windows_open_closed_timing_spectrum} shows the arrival time spectrum for two series with identical run parameters except the blue histogram was taken with the windows closed and the red histogram was taken with the windows open. One can see that for otherwise identical arrival time spectra, there is a higher fraction of \Htwop\ in the configuration with the higher pressure (i.e., Si windows open).

Residual gas analyzer (RGA) data were taken periodically during operations, usually at the end of a reactor cycle. The RGA was positioned in the room-temperature region of the proton detector vacuum section, as illustrated in Fig.~\ref{fig:BL2beamline}. Although these were not direct measurements of the residual gases near the trap, the relative trends in the RGA spectra were useful. In the trap, the partial pressures of species other than hydrogen and helium are expected to be significantly reduced due to cryopumping, as discussed in Section~\ref{sec:ResidualGas}. Regardless, it was possible to track correlations between the residual gas and the timing spectra. For example, when the vacuum system was opened to change a proton detector, the subsequent pump-down and bake-out were not necessarily identical, leading to measurably different pressure conditions.

\begin{figure}[h!]
    \centering
    \includegraphics[width=.45\textwidth]{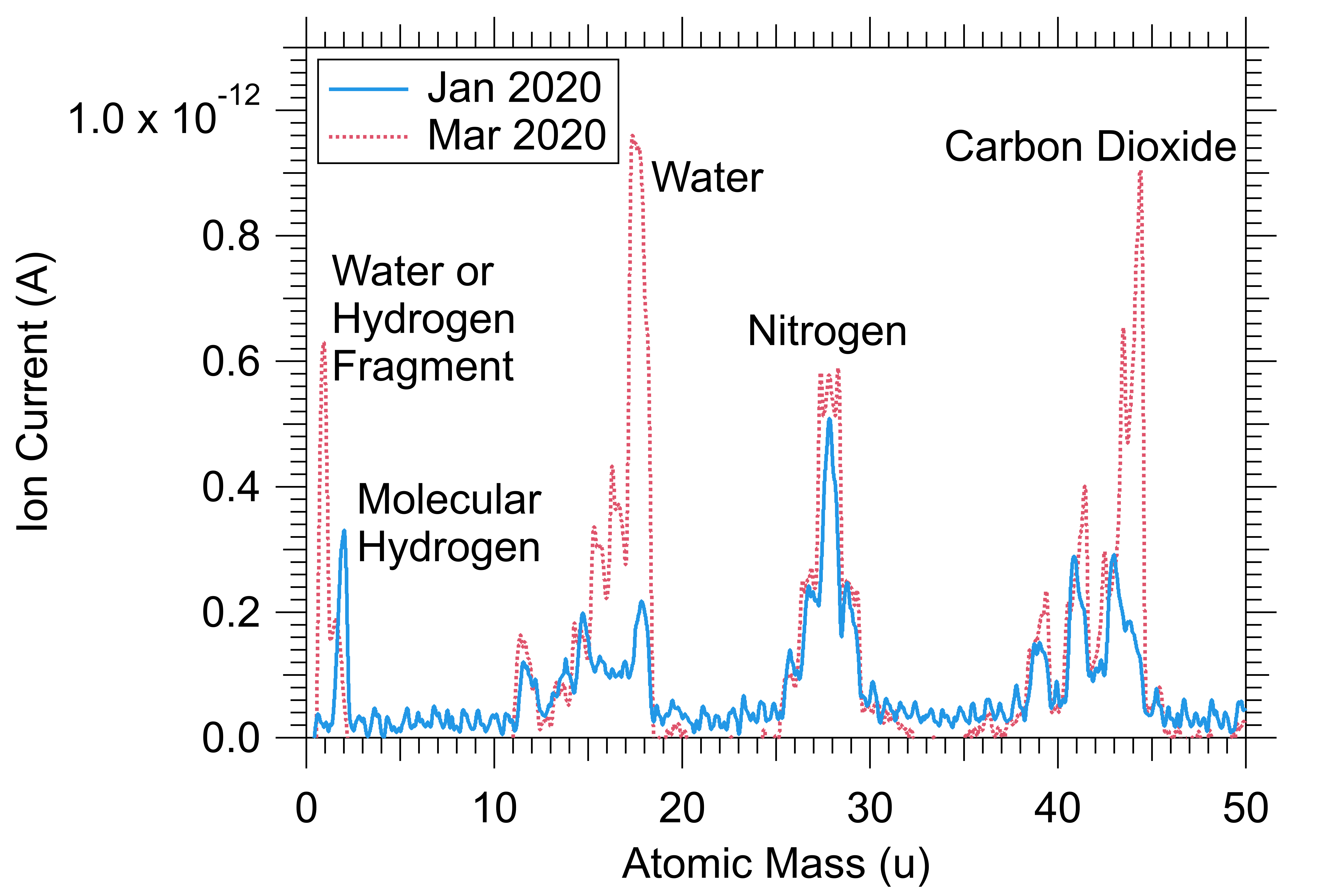}
    \caption{RGA spectra taken in the January and March 2020 reactor cycles.}
    \label{fig:RGA_comp}
\end{figure}

\begin{figure}[h!]
    \centering
    \includegraphics[width=.45\textwidth]{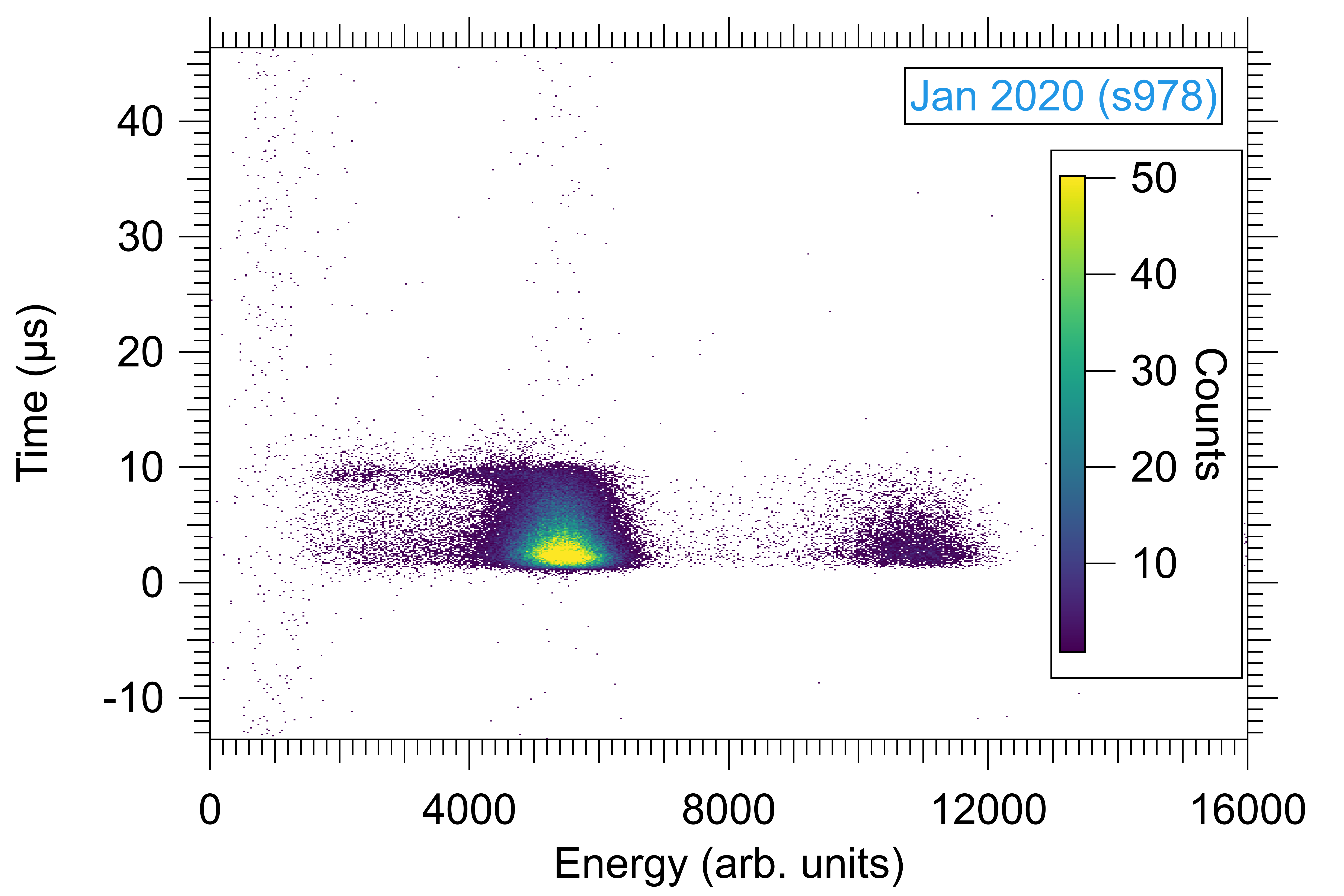}
    \includegraphics[width=.45\textwidth]{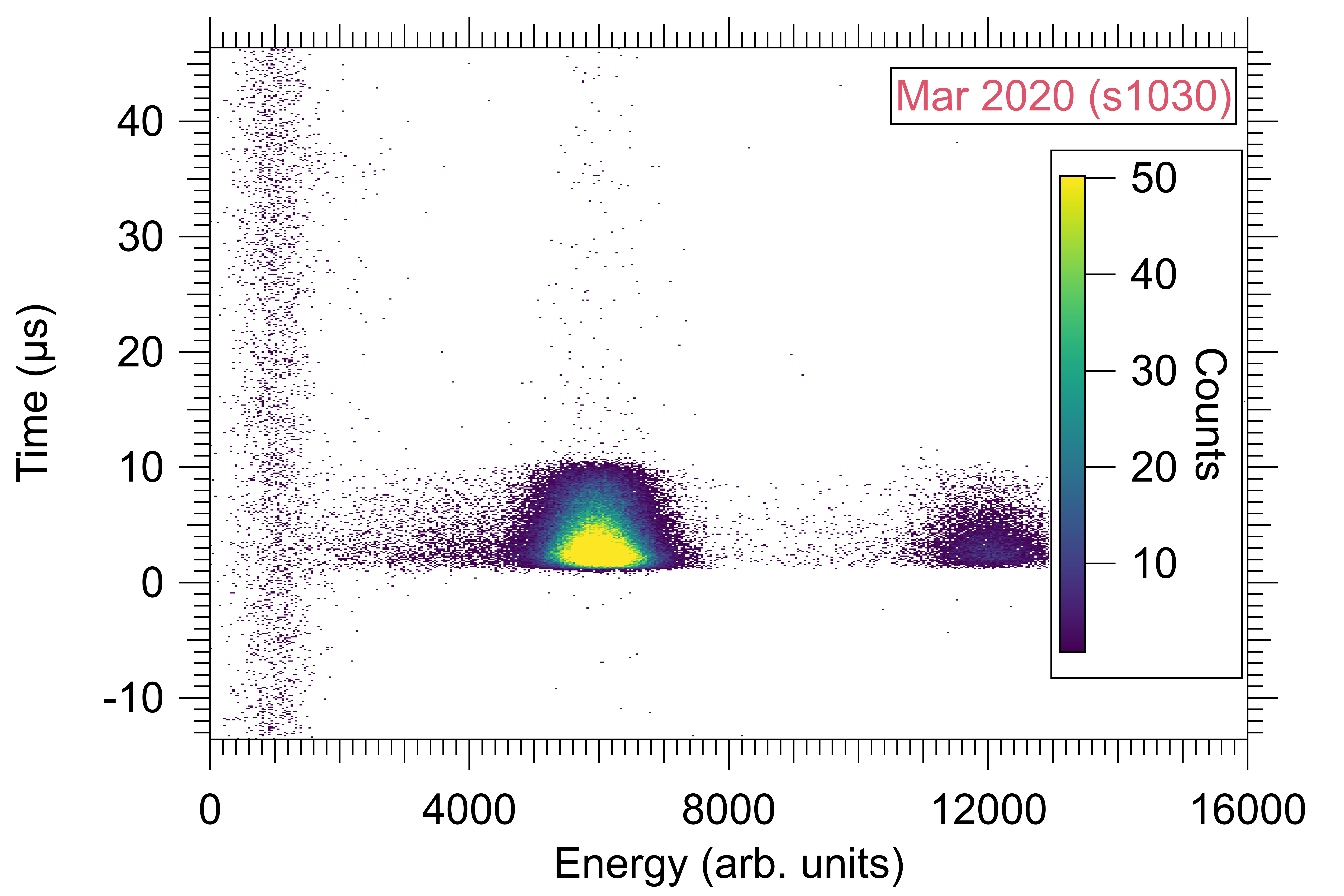}
    \caption{Two data runs with a 45\,V ramp from the January and March 2020 reactor cycles. The top plot is from January, where the RGA spectrum showed an indication of molecular hydrogen in the vacuum system. In contrast, the bottom plot from March 2020 shows no indication of trapped \Htwop\ ions or there being molecular hydrogen in the vacuum system at any significant level.}
    \label{fig:RGA_data}
\end{figure}

Figure~\ref{fig:RGA_comp} shows RGA spectra taken for two different reactor cycles, one in January 2020 and one in March 2020. While the March 2020 cycle had much more \HtwoO\ and \COtwo\ in the vacuum, the January 2020 RGA spectra showed an increased amount of mass 2, indicating the presence of molecular hydrogen. Figure~\ref{fig:RGA_data} plots data taken in the January 2020 and March 2020 reactor cycles showing a significant difference in the region where the \Htwop\ ions appear. The March 2020 cycle shows no indication of an \Htwop\ peak even though the overall pressure from that cycle is higher. The January 2020 cycle shows events in the energy and timing region that are consistent with the trapping of \Htwop\ ions (see Sections~\ref{subsec:ArrivalTime} and \ref{subsec:EnergyMatching}) and are also consistent with the existence of molecular hydrogen in the RGA spectrum.

\begin{figure}[h!]
    \centering
    \includegraphics[width=.45\textwidth]{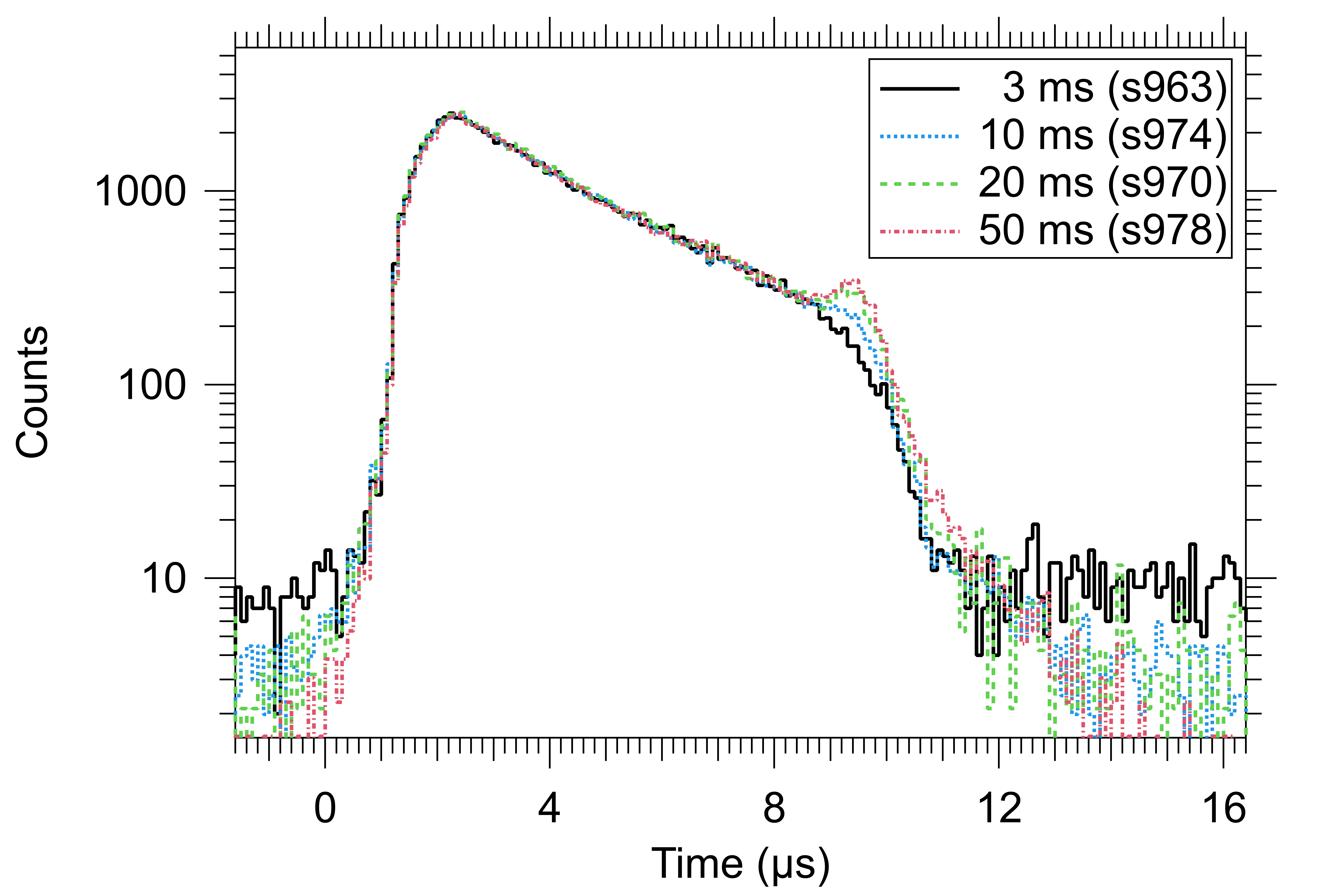}
    \caption{Comparison of the proton timing spectra with the Si windows closed, showing the increase in the \Htwop\ events with increased trapping time. Spectra have different normalizations to allow for a visual comparison. }
    \label{fig:windows_closed_timing_spectra}
\end{figure}

The charge exchange probability is also dependent on the distance that the proton has traveled through the residual gas. Because the decay proton energy distribution doesn't change, the average distance traveled is proportional to the amount of time a trapped proton spends in the trap. Consequently, the \Htwop\ peak fraction should increase with longer trapping time. Figure~\ref{fig:windows_closed_timing_spectra} shows a comparison of timing spectra between four different trapping times with all other experimental parameters kept the same. It is apparent that as the trapping time increases, the \Htwop\ fraction also increases, consistent with expectations. The positive correlation in the data between the two variables, residual gas density and trapping time, supports the idea that charge exchange is the mechanism leading to \Htwop\ in the trap.

\subsection{\label{subsec:ArrivalTime}Timing Simulation of \Htwop\ Trapping and Transport}

To further understand the trapping of \Htwop\ ions, one can simulate the behavior by Monte Carlo modeling. Because the low energy \Htwop\ ions arrival times are largely controlled by the ramp voltage, a comparison to simulation with different ramp voltages should exhibit similar behavior to that seen in the data. In the data, the \Htwop\ ions arrive at similar times with the latest arriving protons and subsequently move later in time with smaller ramp voltages, as seen in Fig.~\ref{fig:arrival_bonus_peak_ramp_voltage}.

Starting positions for beta-decay protons simulated in GEANT4~\cite{Agostinelli2003,Allison2006,Allison2016} were determined from an independent Monte-Carlo code  (MATLAB~\cite{MATLAB}) that determined neutron locations within the trap for the given neutron phase space at NG-C, beam collimation, and effect of gravity.  Those neutron positions were used for proton starting positions within the trap in the GEANT4 simulation. To obtain good agreement with the data, it was essential to incorporate a detailed physical model of the apparatus, including the electric and magnetic fields.

The first step of the GEANT4 simulation is to trap the protons for 30\,$\mu$s with the door and mirror at 800\,V (each 3 electrodes long) to give the protons time to sample the trap volume.  Positions and momenta are stored after this 30\,$\mu$s period to be used in the second step, which transports the protons to the proton detector. The holding time in the simulation is abbreviated compared to the experimental holding time because it is computationally expensive to allow the particles to propagate and the simulated holding time is enough so that all but the lowest energy protons make multiple bounces back and forth in the trap.

\begin{figure}[h!]
  \centering
  \includegraphics[width=.45\textwidth]{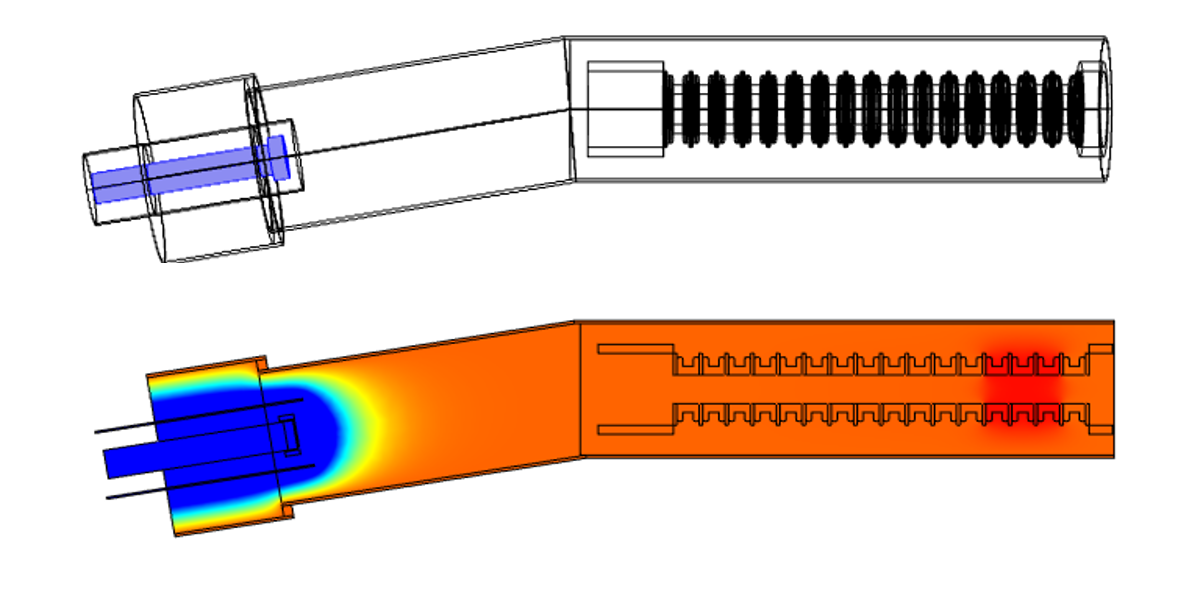}
  \caption{Geometry of trap electrodes and detector used in COMSOL field calculations (top). Electric field profile for 9-electrode length trap (bottom). The dark blue region is where the detector is at high voltage.}
  \label{fig:comsol_fields}
\end{figure}

In the second step, the door is set to a small negative voltage and the appropriate ramp voltages are applied to the trap electrodes.  The three-dimensional electric field maps were created in COMSOL~\cite{COMSOL} for the given electrode geometry and voltages and a detector voltage of $-25$\,kV, as shown in Fig.~\ref{fig:comsol_fields}.  These electric fields and the COMSOL-simulated magnetic field of the superconducting coils are imported into GEANT4. While the proton energy is determined from the beta-decay proton spectrum, deuterons simulating \Htwo\ ions are given an energy of 0.025\,eV.  Any \Htwop\ in the trap would be colder, but 0.025\,eV is already so small that essentially all momentum the ions have as they move toward the detector is from the ramp voltages or the slope of the mirror voltage. The time it takes for each proton and deuteron (\Htwo) to reach the detector position is tallied. The simulation is run for all ramp configurations that were used for data taking.

The 9-electrode data were taken with ramp voltages between 0\,V and 45\,V. In Fig.~\ref{fig:arrival_bonus_peak_ramp_voltage} one can see the main proton peak that arrives soon after the door opens. This peak does not change significantly with a change in the ramp voltage because the majority of beta-decay protons have much more energy than the applied ramp voltage. However, the \Htwop\ peak, seen as the bump between 10\,$\mu$s\ and 22\,$\mu$s after the door opens, does vary significantly with the changing ramp voltage. One can also see that when there is no ramp voltage, the main proton peak arrives at the same time as the other data, but there is no \Htwop\ peak because there is no ramp voltage to bunch the ions in time.

\begin{figure}[h!]
  \centering
  \includegraphics[width=.45\textwidth]{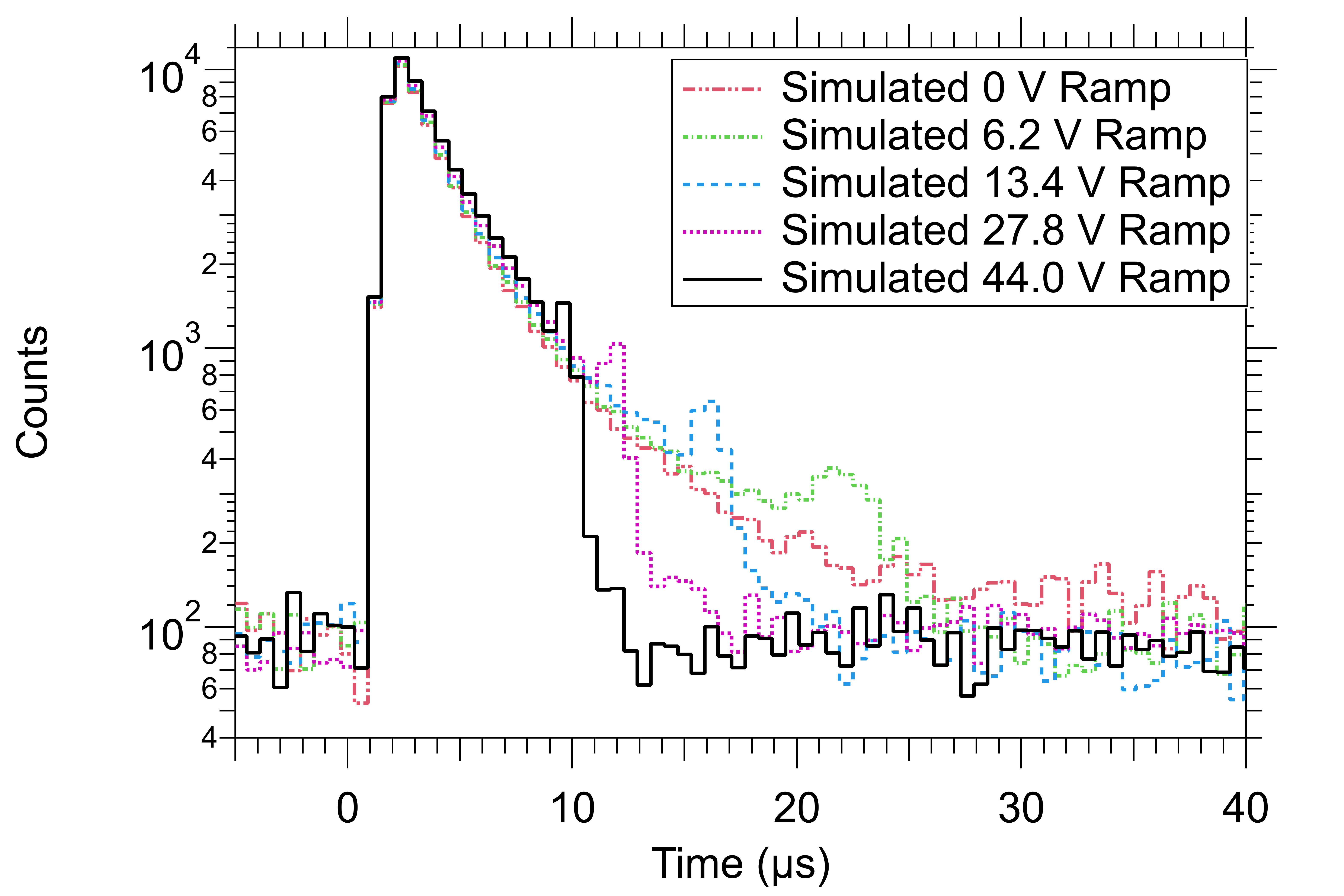}
  \caption{Simulated timing spectra with a 3\,\% contamination of \Htwop. Simulated arrival time of the \Htwop\ peak is in good agreement with the data in Fig.~\ref{fig:arrival_bonus_peak_ramp_voltage}}
  \label{fig:arrival_time_sim}
\end{figure}

An exact comparison between the data and the timing simulation can be made. Figure~\ref{fig:arrival_time_sim} shows the results of the GEANT4 simulation for the same apparatus configuration as the data in Fig.~\ref{fig:arrival_bonus_peak_ramp_voltage}. The simulated timing spectra are generated by combining the arrival times of the protons with a small fraction of the arrival times of the \Htwop\ ions. For comparison, the combined simulated spectra are comprised of 3\,\% \Htwop\ and 97\,\% protons. This is a typical fraction found from the analysis in Section~\ref{subsec:EnergyMatching} and matches well to the data in Fig.~\ref{fig:arrival_bonus_peak_ramp_voltage}. The simulated data qualitatively matches the way the shape of the spectrum varies with ramp voltage, as shown in Fig.~\ref{fig:arrival_bonus_peak_ramp_voltage}. 

The data in Fig.~\ref{fig:arrival_bonus_peak_ramp_voltage} were taken early in the operation of the experiment, and subsequently, it was discovered that the power supply used for the ramp voltage was not designed for precisely supplying such low voltages, and the actual applied voltage may have deviated slightly from the nominal voltage. The simulation shows better agreement with a $-1$\,V offset from the nominal applied voltage, and thus those are the voltages used in Fig~\ref{fig:arrival_time_sim}. The difference in the arrival time of the initial proton peak is unaffected by this shift. However the low energy \Htwop\ ions arrive between 0.5\,$\mu$s and 2\,$\mu$s later with this offset. This does not change the general findings that we see very good agreement between the data and the simulated arrival time when adding a small fraction of very low energy \Htwop\ ions to the data. 

\begin{figure}[h!]
  \centering
  \includegraphics[width=.45\textwidth]{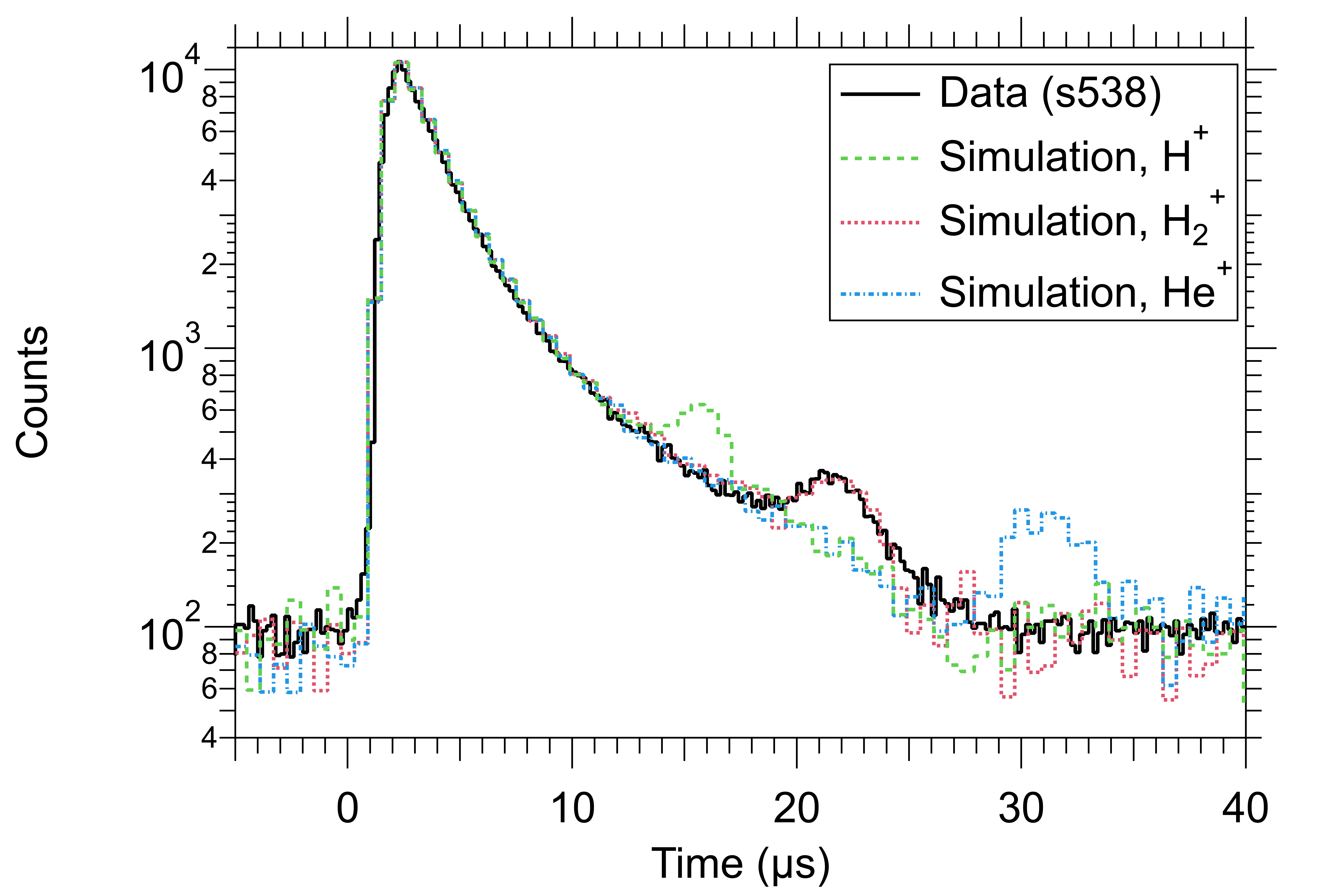}
  \caption{Comparison of data with simulated timing spectra for charged particles of different mass. The simulation of \Htwop\ is consistent with the data while other candidate ions are not.}
  \label{fig:arrival time mass 1, 2 and 4}
\end{figure}

Figure~\ref{fig:arrival time mass 1, 2 and 4} shows 9 electrode data with a nominal ramp voltage of 7.2\,V compared with the simulated combined proton and \Htwop\ timing spectra with a ramp voltage of 6.2\,V with good agreement. The agreement on the width is notable because it depends on an accurate simulation of the geometry.  Also plotted in Fig.~\ref{fig:arrival time mass 1, 2 and 4} are the combined timing spectra for a proton and singly charged helium atom with initial energies of 0.025\,eV, the same as the \Htwop. This shows what the arrival time spectra would look like for other possible light ions. It can be seen that the thermal energy protons and helium arrive earlier and later, respectively, than the peak seen in the data. This is additional evidence that the extra events seen in the data are \Htwop\ ions.

\section{\label{sec:H2Measurement}Numerical Estimations of the Effect of \Htwop\ Ions}

We have shown that it is possible to generate and detect \Htwop\ ions created via charge exchange with a trapped proton under the appropriate conditions. It is also possible to identify those ions experimentally. While one can significantly reduce the number of ions that are generated through careful control of vacuum conditions, it is still important to quantify the effect that a given population of measured \Htwop\ ions would have on the neutron lifetime. In this section, we characterize the energy and backscattering behavior of detected ions and compare the data with simulations to produce quantitative estimates of the effect on a measured lifetime.

\subsection{\label{subsec:EnergyLoss}Energy Loss}

The spectrum of the deposited energy for protons and \Htwop\ ions will differ because of the energy loss in the silicon detector deadlayer. Both SB and PIPS detectors have an inactive layer of gold or silicon dioxide on the surface. Using the difference in energy spectra from several different experimental configurations, a comparison of the \Htwop\ energy spectra can be made to that of the proton spectra. By changing the trapping time, the fraction of \Htwop\ events changes relative to the number of proton events. The \Htwop\ energy can be isolated from the proton energy by subtracting spectra with different trapping times or by isolating the \Htwop\ events in the timing spectra. After the \Htwop\ energy spectra are determined, they may be compared to simulation. 

The \Htwop\ energy spectra were simulated using the SRIM software package~\cite{SRIM_2013,SRIM_book_2015}. SRIM simulations were done for protons at their full acceleration potential (ranging from $-25$\,kV to $-35$\,kV) and for protons with half the acceleration potential. Given the comparatively small \Htwo\ molecular binding energy of 4.8\,eV, we assume that the two protons in the \Htwo\ molecule act independently as they enter the detector. Therefore, one can model \Htwop\ as two independent protons each with half the energy of the acceleration potential. Each of these ``half-energy" protons is then simulated so that the energy loss and backscatter fraction are determined independently.

\begin{figure}[ht]
  \centering
  \includegraphics[width=.45\textwidth]{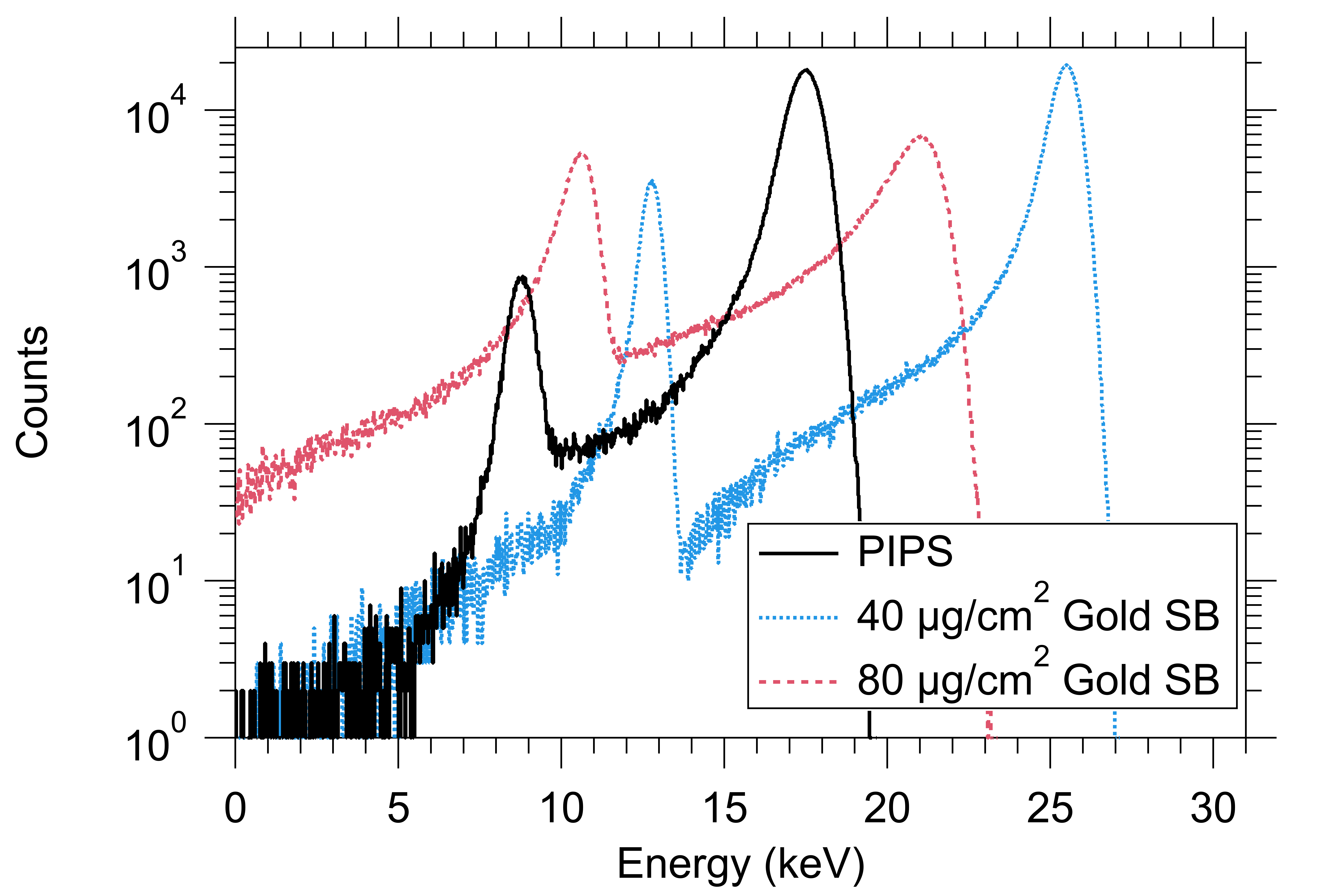}
  \caption{Simulated \Htwop\ energy spectra for three different detectors. The main features are described in the text.}
  \label{fig:H2_half_energy}
\end{figure}

To reconstruct the full \Htwop\ deposited energy, the simulated energy spectra for the half-energy protons are treated as a probability density function. Using this function, a Monte Carlo was written to select two half-energy proton energies for a given acceleration voltage. These two energies are then added together to make a new total energy spectrum for the \Htwop. Figure~\ref{fig:H2_half_energy} shows examples of spectra generated by the SRIM output and the Monte Carlo for a PIPS detector and two SB detectors with different gold thicknesses. The result is an energy spectrum that is slightly wider and has lower energy than the normal proton spectrum. The \Htwop\ spectrum is slightly lower energy because each of the two half-energy protons loses some energy through the deadlayer.

\subsection{\label{subsec:Backscatter}Backscatter Fraction}

Because of the absolute counting nature of the beam neutron lifetime experiments, it is critical to understand proton backscattering from the detector. To account for backscattering in a lifetime measurement, data are taken with different detectors at differing acceleration potentials. The PIPS detectors have more energy loss but the smallest backscatter fraction while SB detectors generally have less energy loss and a larger backscatter fraction. Using multiple detector configurations, data can be taken with a backscatter fraction that varies by an order of magnitude, and the neutron lifetime can be obtained by extrapolating out the backscattering effect. If one were detecting \Htwop\ ions instead of protons, one must quantitatively determine its contribution to the backscattering fraction.

Each proton has energy loss through the deadlayer and a probability of backscattering. A trapped proton has two possible fates after it backscatters from the detector. It can either be re-accelerated by the electric field and return to the detector or backscatter at such an angle that it misses the detector and is completely lost. If the proton re-enters the detector, it may not have enough energy to get completely through the deadlayer, and it may be lost. If the proton has enough energy to get through the deadlayer, it deposits its energy into the active region but with less deposited energy on average than a typical proton.

For \Htwop, each of the two protons has all of these options, leading to three main cases that must be considered: 1) both half-energy protons from \Htwop\ enter the detector on their first hit, which leads to a slightly lower energy peak than a single accelerated trapped proton; 2) both \Htwop\ protons backscatter in such a way that they are both completely lost; or 3) one \Htwop\ proton backscatters in such a way that it is lost and the other enters the detector on its first hit. This causes a unique feature in the \Htwop\ energy spectrum where there are two peaks visible: one at slightly lower energy than the main proton peak, and a second peak at roughly half the energy of the main \Htwop\ peak corresponding to only a single \Htwop\ proton being detected, as seen in Fig.~\ref{fig:H2_half_energy}. As the amount of these \Htwop\ events increases with increased trapping time, this half-energy peak becomes more visible.

\begin{figure}[ht]
    \centering
    \includegraphics[width=.45\textwidth]{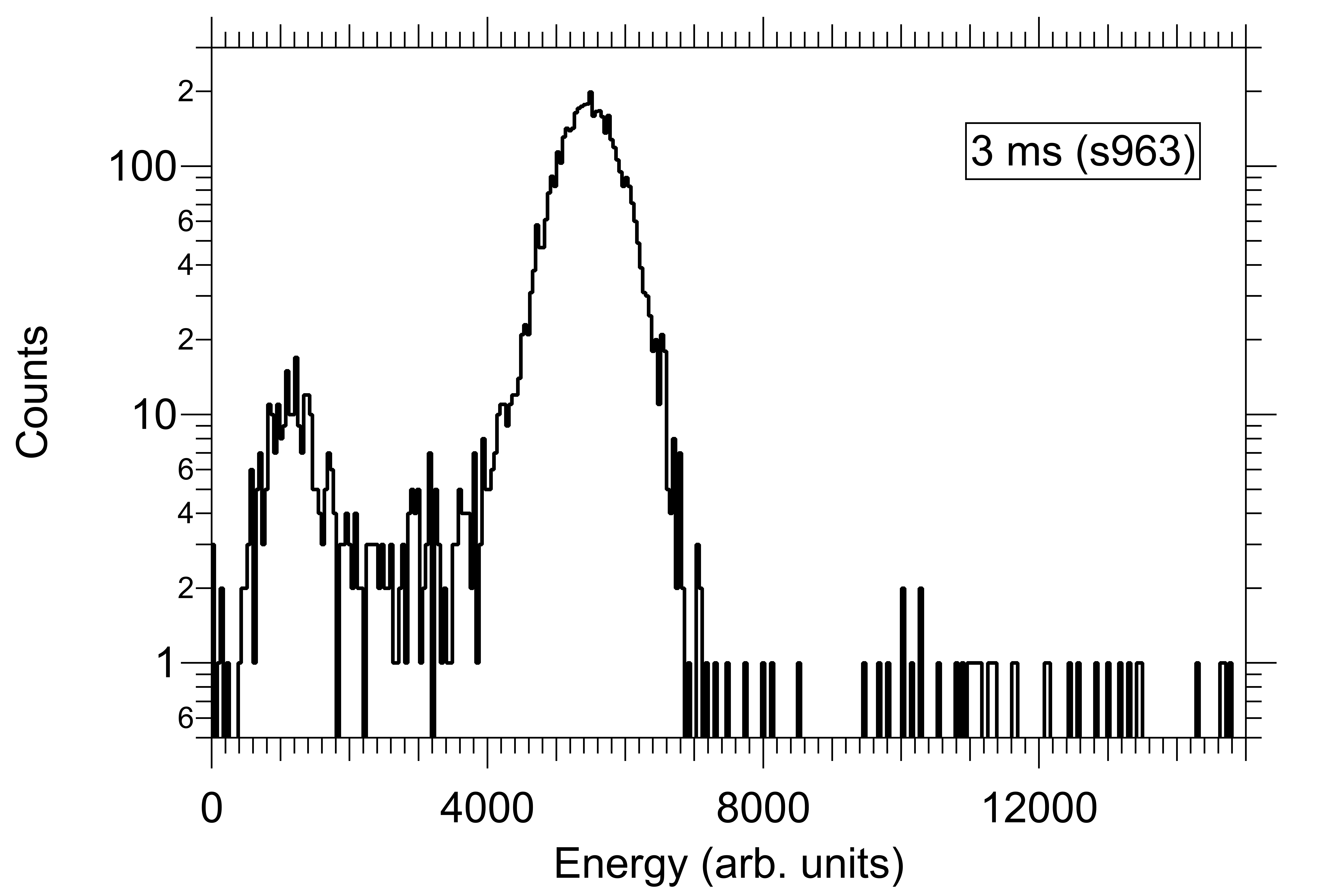}
    \includegraphics[width=.45\textwidth]{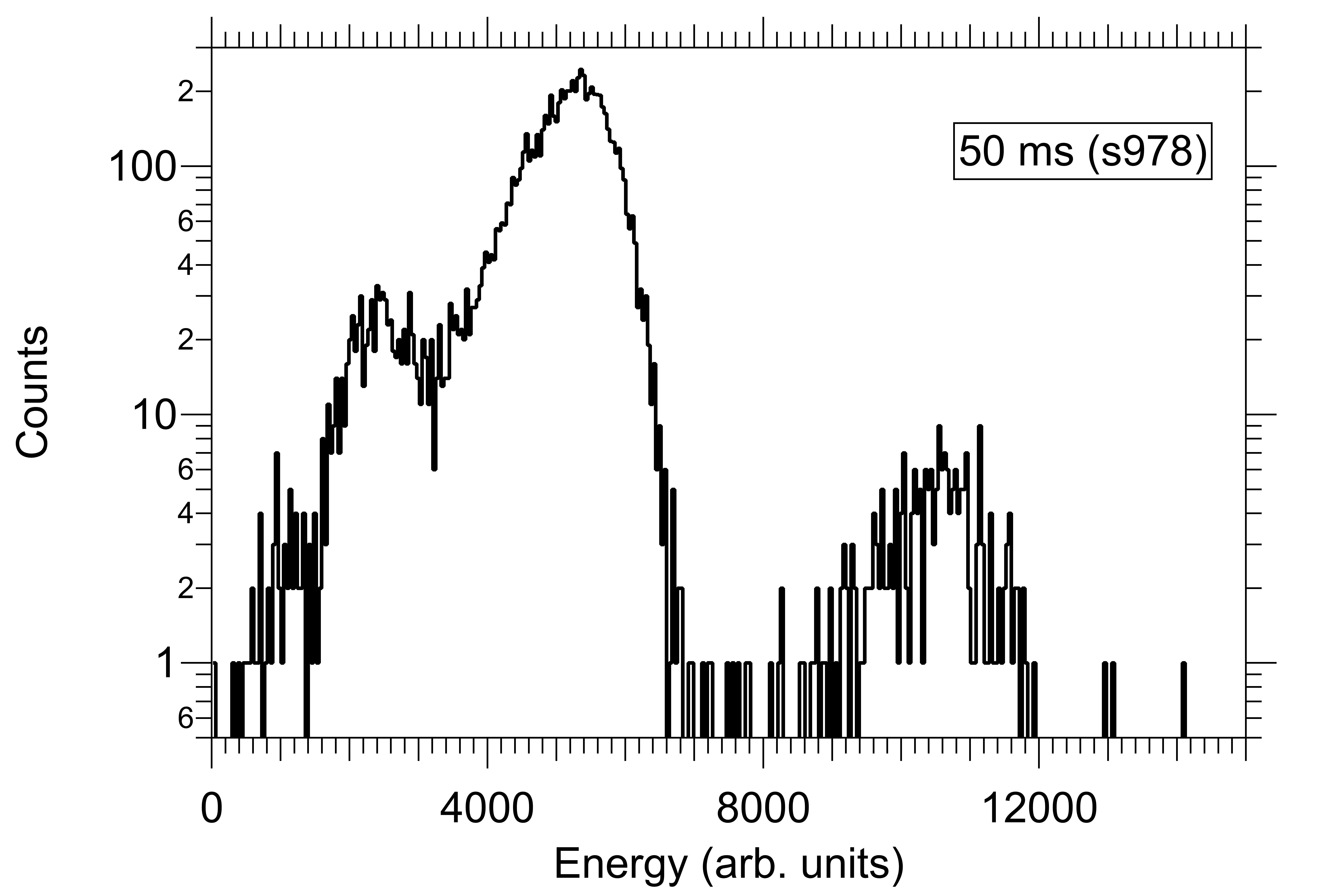}
    \caption{Comparison of energy spectra at 3\,ms (top) and 50\,ms (bottom) showing the increase of backscattered \Htwop\ events with increased trapping time. The change in shape of the main peak comes from an admixture of the slightly lower energy \Htwop\ events. The peak at twice the proton energy (around ch. 10,500) arises from two protons being trapped and is larger in the bottom plot due to the longer trapping time. The trapping of multiple protons is accounted for when determining the total number of protons for a lifetime measurement but is not relevant to this discussion of \Htwo\ contamination. The lowest energy events (below ch. 2000) are predominantly detector noise.}
    \label{fig:energy_spec_backscatter_trap_time}
\end{figure}

Figure~\ref{fig:energy_spec_backscatter_trap_time} shows measured energy spectra for two different trap times (3\,ms and 50\,ms) with timing cuts made to reduce random background events and enhance the visibility of the \Htwop\ peak. Both spectra used the same detector and were taken using a 45 V ramp and a trap length of 9 electrodes. As the trapping time increases, the half-energy peak becomes clearly visible. The Monte Carlo used to generate the \Htwop\ energy spectrum takes into account the probability of each of the half-energy protons to backscatter or be stopped in the deadlayer. The SRIM + Monte Carlo \Htwop\ energy spectrum encapsulates this, resulting in a main \Htwop\ peak, slightly lower in energy than the normal proton peak, and a half-energy peak that is caused by a single half-energy proton, as seen in Fig.~\ref{fig:H2_half_energy}. While this simulation does not produce an estimate of the dependence of the half-energy peak on trapping time, Fig.~\ref{fig:H2_half_energy} shows that the simulated half-energy peak behaves as expected, increasing in amplitude as the backscatter fraction of the detector increases.

\subsection{\label{subsec:EnergyMatching}Simulation of Energy Spectra with \Htwop\ Ions}

The discussions thus far of the effect of \Htwop\ ions in proton counting data have been qualitative, but it is important to understand quantitatively how \Htwop\ events from charge exchange could affect the absolute determination of the number of protons from neutron decay. Because the \Htwop\ ion arrives at the proton detector in a time range overlapping with the trapped protons, the energy spectra of the two would be difficult to disentangle. The measured energy spectrum, however, will be some combination of the proton energy spectrum and the \Htwop\ energy spectrum, and so it is possible to simulate the effect and scale it to match it to the data.

Using SRIM + Monte Carlo, the \Htwop\ energy spectra and the normal proton spectra can be compared to the measured data. If \Htwop\ were present in the trap, some admixture of the \Htwop\ and proton spectra would best fit the data. Two test cases are examined here: one with a negligible amount of \Htwop\ visible (PIPS), and one with a large amount of \Htwop\ visible (100\,\microgcm\ SB detector). To do this, noise must be added to the simulated spectra to match the width of the data. Approximately 1.3\,keV of Gaussian noise is added to both the \Htwop\ and the proton spectra. The same 1.3\,keV of added noise matches well with most of the data, suggesting that the noise is stable over long periods of running. After the noise is added, the simulated spectra may be compared directly to the data.

\begin{figure}[ht]
  \centering
  \includegraphics[width=.45\textwidth]{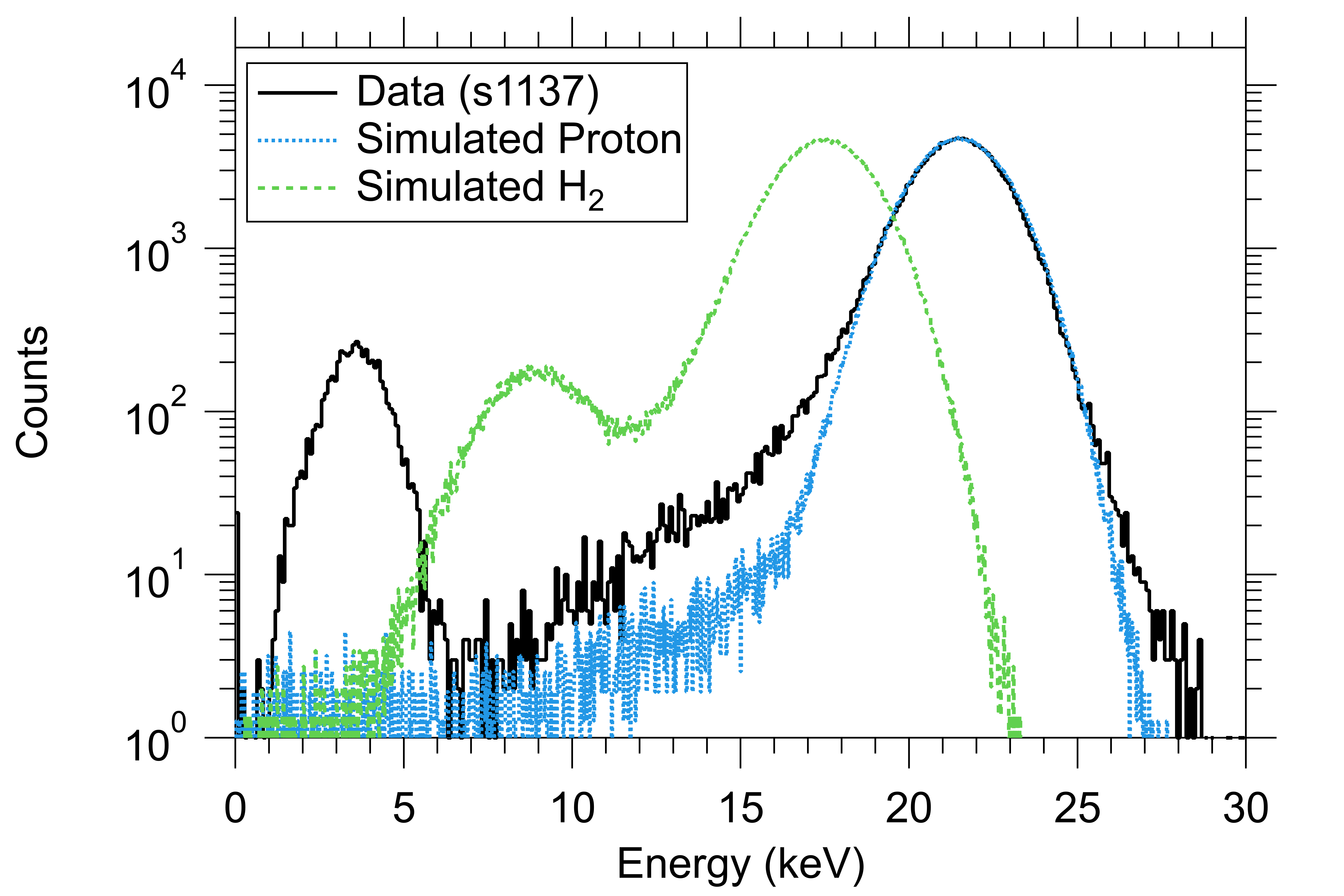}
  \includegraphics[width=.45\textwidth]{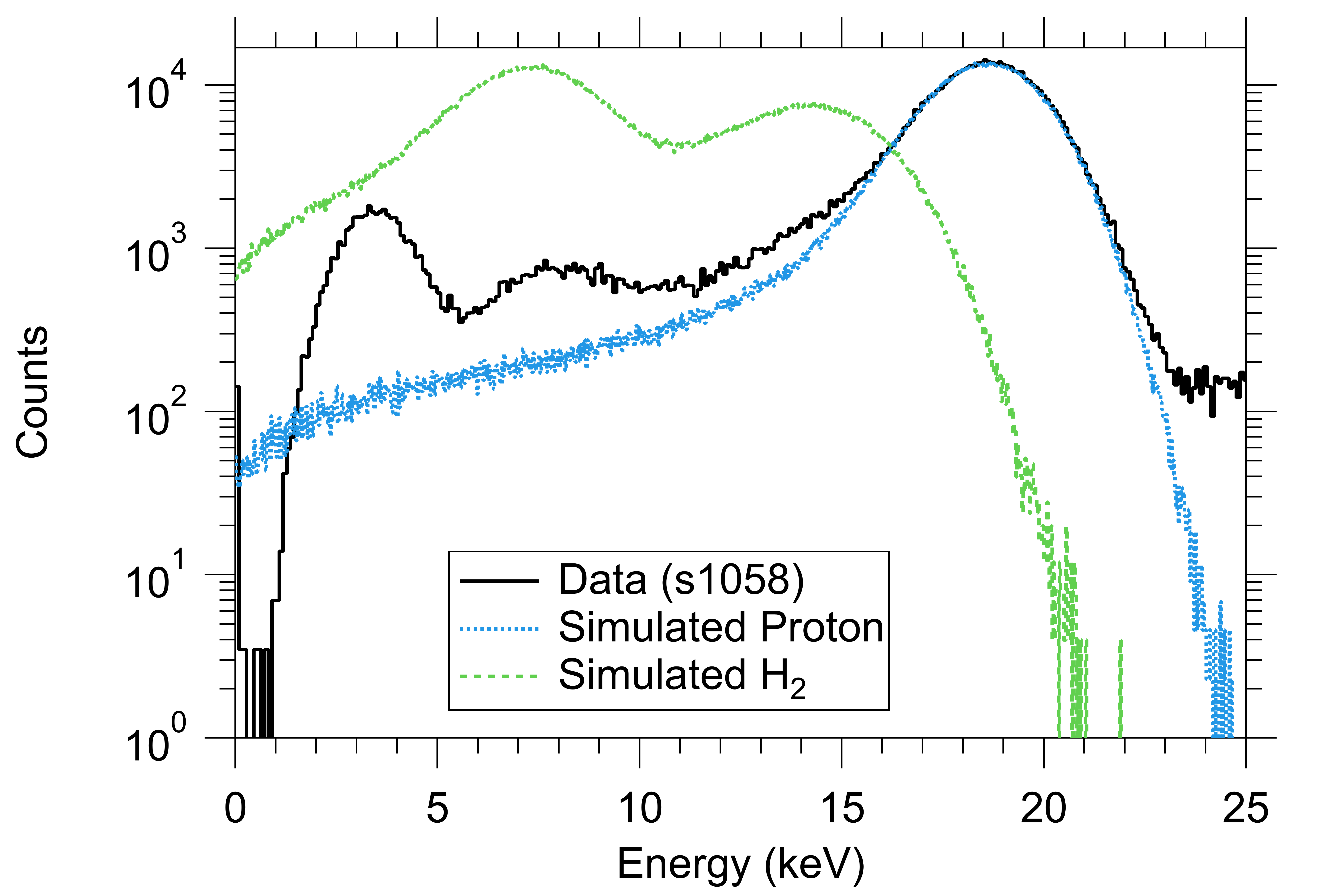}
  \caption{Top: Comparison of PIPS detector data (black) with the SRIM simulated energy spectra of the proton (blue) and \Htwop\ (green). Bottom: Comparison of 100\,\microgcm\ SB detector data (black) with the SRIM simulated energy spectra of the proton (blue) and \Htwop\ (green). Both plots illustrate that the proton spectrum alone is not sufficient to match the data. The lowest energy peak in the data in both plots corresponds to the noise threshold.}
  \label{fig:H_H2_data_comp}
\end{figure}

Figure~\ref{fig:H_H2_data_comp} shows data for the two cases along with simulations of the proton spectrum and the \Htwop\ spectrum. In both cases, the data were acquired with a 45 V ramp and a trap length of 9 electrodes. The simulated proton spectrum agrees with the data in the central peak region and the high-energy shoulder of the peak, but it starts to deviate from the data on the low-energy shoulder and the low-energy tail. It is clear that the \Htwop\ spectrum itself does not match the data well, suggesting that if \Htwo\ were present in the proton trap, it makes up a relatively small fraction of the total events.

Comparing the upper and lower panels of Fig.~\ref{fig:H_H2_data_comp} shows that the half-energy peak of the \Htwop\ spectra varies significantly depending on the detector type. Because the half-energy peak is caused by a single backscattered half-energy proton, the probability of a half-energy event increases when the backscatter fraction of the half-energy proton increases. The PIPS detector, with its small backscatter fraction, produces a significantly smaller half-energy peak. The 100\,\microgcm\ SB detector has the highest backscatter fraction of any detector used in these measurements. With this detector, the probability that a half-energy proton does not pass into the active region without backscattering is \(> 30\,\%\); thus, the size of the half-energy peak is larger than that of the full energy \Htwo\ peak. 

\begin{figure}[ht]
  \centering
  \includegraphics[width=.45\textwidth]{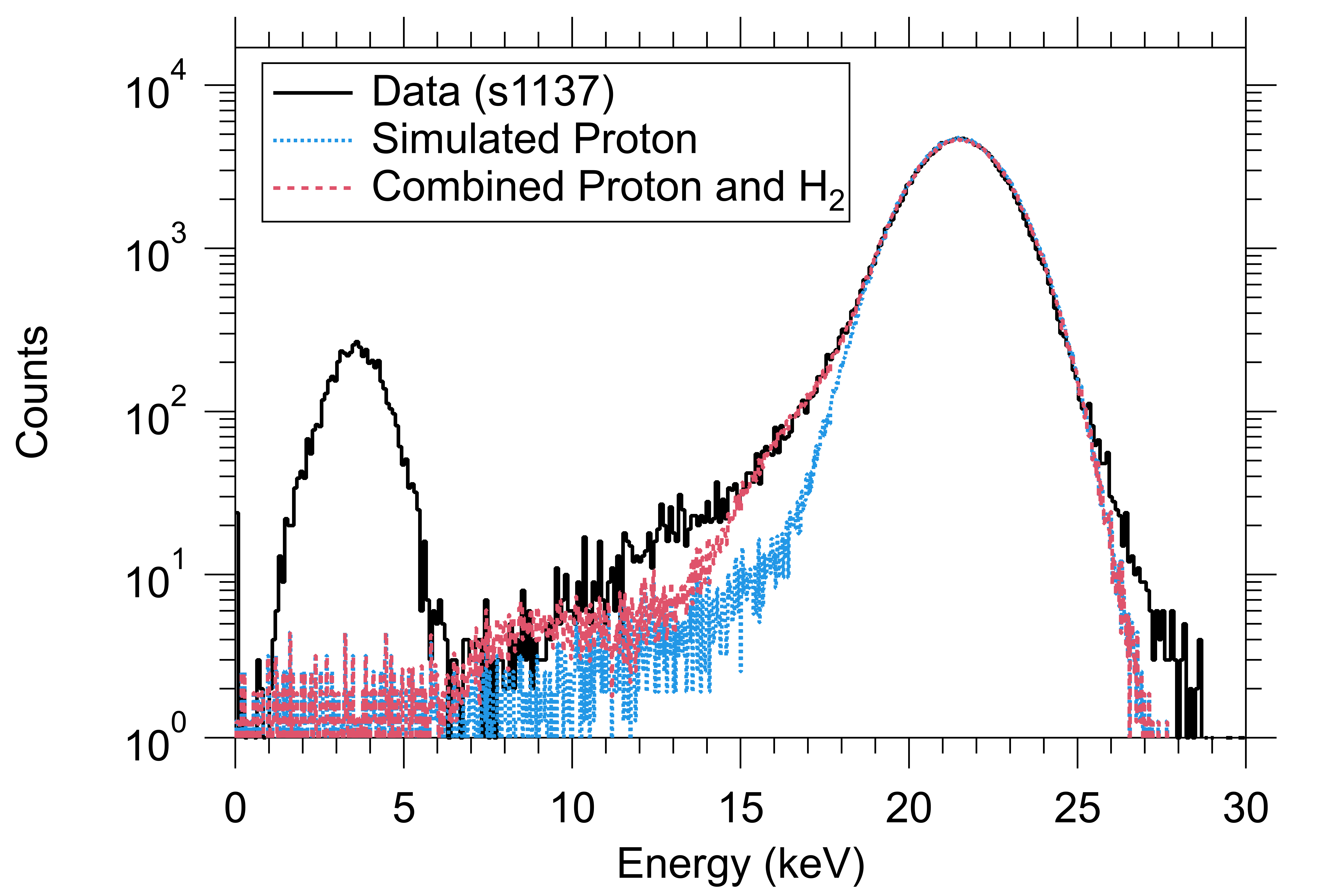}
  \includegraphics[width=.45\textwidth]{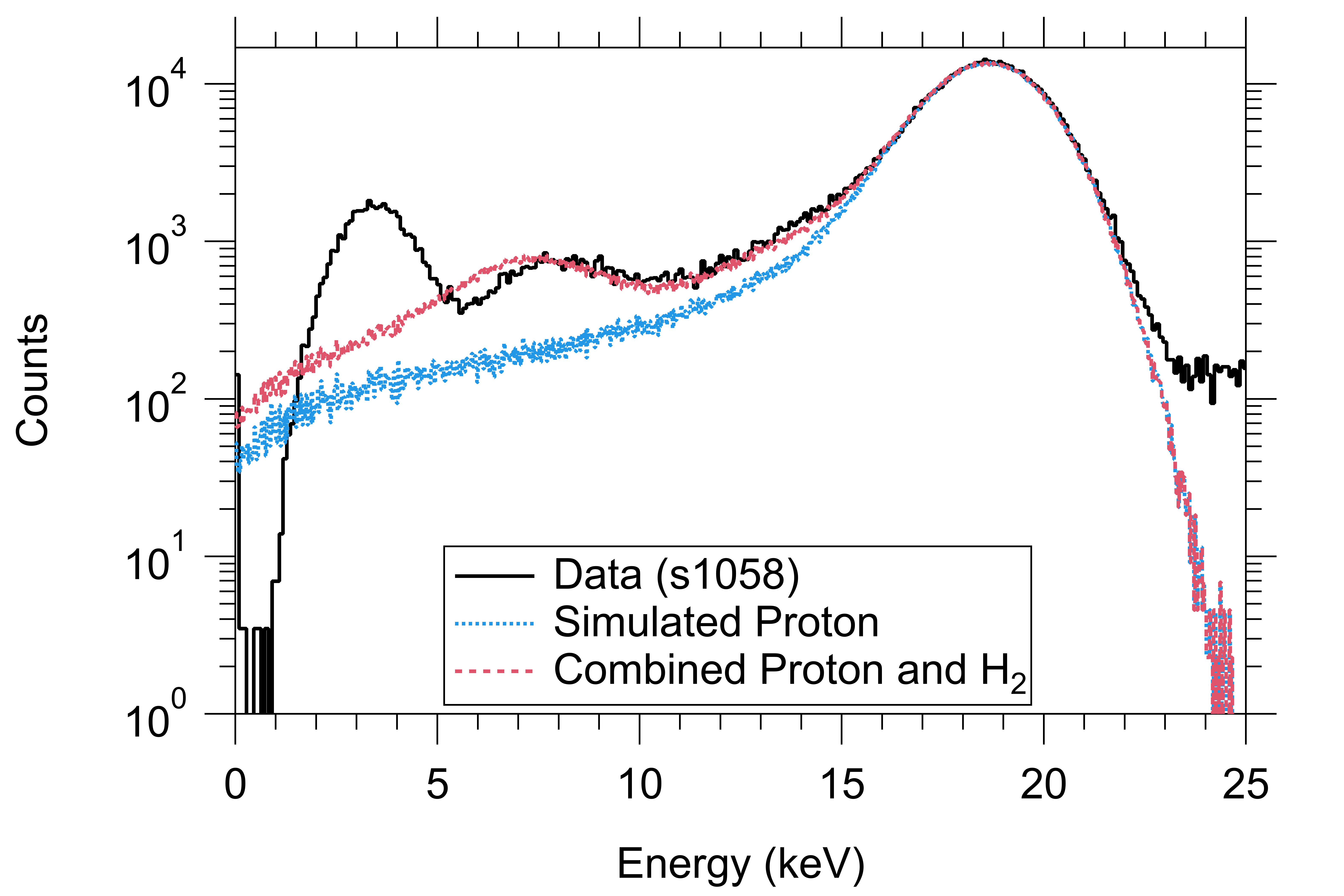}
  \caption{Top: Comparison of PIPS detector data (black) with the SRIM simulated energy spectra of the proton (blue) and combined admixture of the proton and 2\,\% \Htwop\ spectra (red). Bottom: Comparison of 100\,\microgcm\ SB detector data (black) with the SRIM simulated energy spectra of the proton (blue) and combined admixture of the proton and 5\,\% \Htwop\ spectra (red). Both plots illustrate that combined spectra fit the data significantly better than the proton spectra alone.}
  \label{fig:combined_data_comp}
\end{figure}

Using the simulated proton and \Htwop\ energy spectra, an admixture of the two can now be generated to mimic the data with a single free parameter. This combined energy spectrum is produced by taking a small amount of the \Htwop\ spectrum and adding it to the normal proton spectrum until the best match to the data is achieved. This is done independently for each series. As seen in Fig.~\ref{fig:combined_data_comp}, the combined energy spectra fit the data much better than the proton spectrum alone. This indicates that for these two cases, there is some \Htwo\ in the proton trap, and the \Htwop\ is easily detected because of the accelerating potential. The total amount of \Htwop\ in the combined spectrum admixture is about 2\,\% for the PIPS detector comparison and 5\,\% for the SB detector comparison.

While it was asserted that the PIPS detector data for this exercise had a ``negligible amount of \Htwop\ peak visible," this analysis suggests that there is only a factor of 2.5 less \Htwop\ in these data than the SB detector data. During standard diagnostic checks on the data, the tell-tale sign of \Htwop\ in the data is the half-energy peak. Because the backscatter fraction is significantly smaller with the PIPS detector, the half-energy peak is highly suppressed compared to the full energy \Htwop\ peak. This means that even when the half-energy peak is not easily visible that does not mean there is no \Htwop\ present in the data. The timing spectra and a more careful comparison of the spectral shapes, like the examples in Fig.~\ref{fig:combined_data_comp}, are necessary to determine whether or not there is \Htwop\ in the data.

\subsection{\label{subsec:DetEff}Determination of \Htwop\ Detection Efficiency}

The ultimate goal in studying \Htwo\ is to establish quantitatively how the charge exchange of protons with \Htwo\ molecules in the trap would affect the total number of events that are counted. One must determine the detection efficiency for trapped \Htwop\ ions for a wide range of detector configurations.  This procedure is done following the same method for correcting the proton rate as in Ref.~\cite{Nico_2005}. We use the simulation to calculate \(f_{\rm Ruth}\), the Rutherford backscatter fraction; \(f_{\rm Stp}\), the fraction of events stopped in the deadlayer; and \(f_{\rm BT}\), the events lost below the energy threshold. The detection efficiency for protons is given by 

\begin{equation}\label{eqn:protoneff}
	\epsilon_p = 1 - f_{\rm Ruth}({\rm p}) - f_{\rm Stp}({\rm p}) - f_{\rm BT}({\rm p}),
\end{equation}

\noindent where there are three possibilities for a proton to go undetected. In Ref.~\cite{Nico_2005}, there is also a fraction for the protons that backscatter off the active region of the detector and do not deposit energy above the threshold, \(f_{\rm AT}\). Determining \(f_{\rm AT}\) requires doubling the number of simulations, and it gives a negligible change to the neutron lifetime, so it has been excluded here. Because \Htwop\ is modeled as two independent protons, the calculation of Eq.~\ref{eqn:protoneff} is not valid.

There are three possibilities for each of the \Htwop\ protons to go undetected, giving a total of nine distinct outcomes for the two protons. The two protons are identical so the probability for these events is the same, simplifying the calculation. Nevertheless, cross-terms exist where the two protons have different outcomes. Consider the example of an \Htwop\ ion where one of the protons backscatters but the other deposits its full energy in the active region. This event would only go undetected if the deposited energy were below threshold. The complete list of \Htwop\ ion possibilities is given in Table~\ref{table:H2_outcomes}.

\begin{table*}[t]
\caption{Nine possible outcomes for a \Htwop\ ion to be undetected. Events that enter the active region may be undetected if they are below threshold.}
\begin{ruledtabular}
\begin{tabular}{l | c c c} 
Proton outcome  		& Backscattered 			& Stopped in deadlayer		& Enters active region\\ [0.5ex]
 \hline
Backscattered 			& \(f_{\rm Ruth}f_{\rm Ruth}\)	& \(f_{\rm Stp}f_{\rm Ruth}\)	& \(f_{\rm X}f_{\rm Ruth}\)\\
Stopped in deadlayer 	& \(f_{\rm Ruth}f_{\rm Stp}\)	& \(f_{\rm Stp}f_{\rm Stp}\)	& \(f_{\rm X}f_{\rm Stp}\)\\
Enters active region		& \(f_{\rm Ruth}f_{\rm X}\)	& \(f_{\rm Stp}f_{\rm X}\)		& \(f_{\rm X}f_{\rm X}\)\\
\end{tabular}
\end{ruledtabular}
\label{table:H2_outcomes}
\end{table*}

\begin{figure}[t]
  \centering
  \includegraphics[width=.45\textwidth]{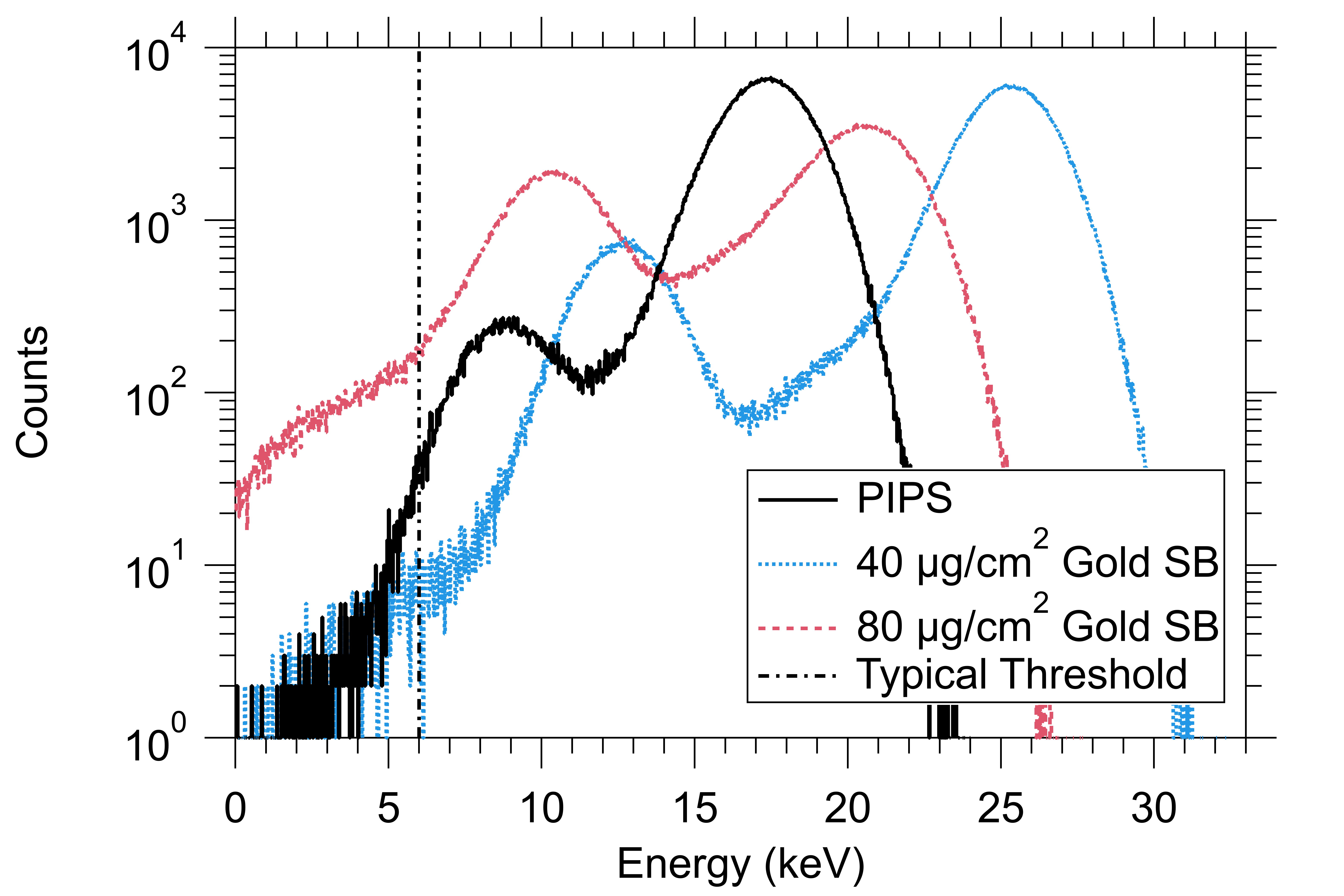}
  \caption{Example of simulated \Htwop\ energy spectra for three different detector types, showing a typical analysis threshold.}
  \label{fig:proton_threshold}
\end{figure}

When adding the two half-energy protons together with the Monte Carlo, the probability for a single proton to backscatter or be stopped in the deadlayer must be taken into account. Therefore, the five outcomes in Table~\ref{table:H2_outcomes} that include the fraction of transmitted events, \(f_{\rm X}(\rm{H_2})\), are consolidated by looking at the simulated \Htwop\ energy spectrum. An example of the typical proton threshold used in analysis overlaid on simulated \Htwop\ spectra can be seen in Fig~\ref{fig:proton_threshold}. \(f_{\rm BT}(\rm{H_2})\) is calculated by taking the fraction of events to the left of the threshold (marked by the dashed line) seen in Fig.~\ref{fig:proton_threshold} and dividing it by the total number of simulated events. As a result, the detection efficiency for an \Htwop\ ion is given by

\begin{align}
\epsilon_{\rm{H_2}} = 1 	&- f_{\rm Ruth}({\rm{H_2}}) - f_{\rm Stp}({\rm{H_2}}) \\
					&- 2(f_{\rm Ruth}({\rm{H}})f_{\rm Stp}({\rm{H}})) - f_{\rm BT}(\rm{H_2}), \nonumber
\end{align}

\noindent where \(f_{\rm Ruth}(\rm{H_2})\) is the fraction of \Htwop\ events that have both half-energy protons backscatter, \(f_{\rm Stp}(\rm{H_2})\) is the fraction of \Htwop\ events that have both half-energy protons stop in the deadlayer, \(f_{\rm Ruth}(\rm{H})\) is the fraction of half-energy events that backscatter, \(f_{\rm Stp}(\rm{H})\) is the fraction of half-energy events that stop in the deadlayer, and \(f_{\rm BT}(\rm{H_2})\) is the fraction of \Htwop\ events that are below threshold. This method has been used to calculate \(\epsilon_{\rm{H_2}}\) for commonly used detector and acceleration voltage configurations.  A summary of the fractions and the detection efficiencies for the proton and the \Htwop\ ion are given in Tables~\ref{table:TRIM_proton_full} and \ref{table:TRIM_H2}, respectively. For the PIPS detector and SB detectors with gold layers of 60\,\microgcm\ and below, the difference in the detection efficiencies between the proton and \Htwop\ is \(< 1.5\,\%\). It is not until the gold deadlayer becomes relatively thick ($\geq 80$\,\microgcm) and the backscatter fraction large ($\gtrapprox 1.75\,\%$) that there are large differences in the detection efficiency.

\begin{table*}[ht]
\caption{To the left of the vertical line are SRIM input parameters for the proton simulation. $\rho_{\rm{Au}}$ is the areal density of the gold layer, and $E_{\rm{H}}$ is the input energy of the proton. To the right of the vertical line are the results for the fractions, $f$, and the determination of the proton detection efficiencies, $\epsilon_{\rm p}$. The values use an energy detection threshold of 6.0\,keV, which is a typical value for the data acquired for these studies.}
\begin{ruledtabular}
\begin{tabular}{c c | c c c r} 
 $\rho_{\rm{Au}}$	& \(E_{\rm{H}}\)	& \(f_{\rm Ruth}\) 	& \(f_{\rm BT}\)	& \(f_{\rm Stp}\)	& \(\epsilon_{\rm p}\)		\\ 	[0.5ex]
 ($\mu$g/cm$^2$) 	& (keV) 		& (\%) 		& (\%) 		& (\%)  		& (\%)				\\ 	[0.5ex]
 \hline
 0 				& 25.0 			& 0.274 		& 0.095		& 0.102 		& 99.5	\\
 0 				& 27.5 		& 0.222 		& 0.066		& 0.067 		& 99.6	\\
  0 				& 30 .0		& 0.189 		& 0.043		& 0.049 		& 99.7	\\
  0 				& 32.5 		& 0.159 		& 0.032		& 0.036 		& 99.8	\\
  0 				& 35.0		& 0.132 		& 0.024		& 0.025 		& 99.8	\\
 20				& 25.0 		& 0.831 		& 0.001		& 0.000 		& 99.2	\\
 20				& 27.5 		& 0.691 		& 0.000		& 0.000 		& 99.3	\\
 20				& 30.0		& 0.589 		& 0.000		& 0.000 		& 99.4	\\
  20				& 32.5 		& 0.518 		& 0.000		& 0.000 		& 99.5	\\
 20				& 35.0		& 0.444 		& 0.000		& 0000		& 99.6	\\
 40				& 25.0		& 1.957 		& 0.031		& 0.002 		& 98.0	\\
 40				& 27.5 		& 1.621 		& 0.017		& 0.000 		& 98.4	\\
 40				& 30.0		& 1.390 		& 0.011		& 0.001 		& 98.6	\\
 40				& 32.5 		& 1.167 		& 0.006		& 0.000 		& 98.8	\\
 40				& 35.0		& 1.025 		& 0.003		& 0.000 		& 99.0	\\
  60				& 25.0		& 3.461 		& 0.207		& 0.043 		& 96.3	\\
 60				& 27.5 		& 2.852 		& 0.121		& 0.027 		& 97.0	\\
 60				& 30.0		& 2.373 		& 0.072		& 0.017 		& 97.5	\\
 60				& 32.5 		& 2.027 		& 0.047		& 0.012 		& 97.9	\\
 60				& 35.0		& 1.749 		& 0.031		& 0.008 		& 98.2	\\
 80				& 25.0		& 5.251 		& 0.602		& 0.319 		& 93.8	\\
 80				& 27.5 		& 4.336 		& 0.365		& 0.203 		& 95.1	\\
 80				& 30.0		& 3.640 		& 0.245		& 0.139 		& 96.0	\\
 80				& 32.5 		& 3.098 		& 0.169		& 0.083 		& 96.7	\\
 80				& 35.0		& 2.655 		& 0.111		& 0.058 		& 97.2	\\
100				& 25.0		& 7.056 		& 1.302		& 1.173 		& 90.5	\\
100				& 27.5 		& 5.890 		& 0.854		& 0.775 		& 92.5	\\
100				& 30.0		& 4.969 		& 0.571		& 0.532 		& 93.9	\\
100				& 32.5 		& 4.255 		& 0.391		& 0.378 		& 95.0	\\
 100				& 35.0		& 3.603 		& 0.285		& 0.253 		& 95.9	\\
\end{tabular}
\end{ruledtabular}
\label{table:TRIM_proton_full}
\end{table*}

\begin{table*}[ht]

\caption{To the left of the vertical line are SRIM input parameters for the half-energy \(\rm{H}^+\) simulations used to determine the detection efficiencies for \Htwop. $\rho_{\rm{Au}}$ is the areal density of the gold layer, and $E_{\rm{H}}$ is the input energy of the proton. To the right of the vertical line are the results for the fractions, $f$, and the determination of the \Htwop\ detection efficiencies, $\epsilon_{H_2}$ using the Monte Carlo method described in the text. The values use an energy detection threshold of 6.0\,keV.}
\begin{ruledtabular}
\begin{tabular}{c c | c c c c r} 
 $\rho_{\rm{Au}}$	& \(E_{\rm{H}}\)	& \(f_{\rm Ruth}(\rm H_2)\)	& \(f_{\rm BT}(\rm H_2)\)	& \(f_{\rm Stp}(\rm H_2)\)	& \(2(f_{\rm Ruth}(\rm H)f_{\rm Stp}(\rm H))\)	& \(\epsilon_{\rm H_2}\)	\\ [0.5ex]
 ($\mu$g/cm$^2$) 	& (keV) 		& (\%) 				& (\%)			& (\%) 				& (\%) 						& (\%) 				\\ [0.5ex]
 \hline
0 				& 12.50 		& 0.013 				& 1.795 			& 0.019 				& 0.031	 					&  98.1	\\
0 				& 13.75 		& 0.009 				& 0.525 			& 0.009 				& 0.018 						&  99.4	\\
0 				& 15.00 		& 0.007 				& 0.133 			& 0.005 				& 0.011 						&  99.9	\\
0 				& 16.25 		& 0.005 				& 0.039 			& 0.002 				& 0.007 						&  100.0	\\
0				& 17.50 		& 0.004 				& 0.016 			& 0.001 				& 0.004 						&  100.0	\\
20				& 12.50 		& 0.078 				& 0.009 			& 0.000 				& 0.000 						&  99.9	\\
20				& 13.75		& 0.058 				& 0.004 			& 0.000 				& 0.000 						&  99.9	\\
20				& 15.00 		& 0.042 				& 0.001 			& 0.000 				& 0.000 						&  100.0	\\
20				& 16.25 		& 0.031 				& 0.001 			& 0.000	 			& 0.000 						&  100.0	\\
20				& 17.50 		& 0.024 				& 0.000			& 0.000 				& 0.000 						&  100.0	\\
40				& 12.50 		& 0.496 				& 0.277 			& 0.000 				& 0.012 						&  99.2	\\
40				& 13.75 		& 0.360 				& 0.136 			& 0.000 				& 0.006 						&  99.5	\\
40				& 15.00 		& 0.264 				& 0.072			& 0.000 				& 0.003 						&  99.7	\\
40				& 16.25 		& 0.200 				& 0.038 			& 0.000 				& 0.002 						& 99.8	\\
40				& 17.50 		& 0.157 				& 0.024 			& 0.000 				& 0.001 						&  99.8	\\
60				& 12.50 		& 1.436 				& 1.934 			& 0.018 				& 0.318 						&  96.3	\\
60				& 13.75 		& 1.076 				& 0.928 			& 0.008 				& 0.180 						&  97.8	\\
60				& 15.00 		& 0.802 				& 0.502 			& 0.003 				& 0.102 						&  98.6	\\
60				& 16.25 		& 0.608 				& 0.299 			& 0.002 				& 0.062 						&  99.0	\\
60				& 17.50 		& 0.472 				& 0.180 			& 0.001 				& 0.039 						&  99.3	\\
80				& 12.50		& 2.520 				& 7.779 			& 0.327 				& 1.816 						&  87.6	\\
80				& 13.75 		& 2.001 				& 3.703 			& 0.167 				& 1.156 						&  93.0	\\
80				& 15.00 		& 1.578 				& 2.001 			& 0.088 				& 0.744 						&  95.6	\\
80				& 16.25 		& 1.252 				& 1.182 			& 0.046 				& 0.479 						&  97.0	\\
80				& 17.50 		& 1.006 				& 0.722 			& 0.026 				& 0.323 						&  97.9	\\
100				& 12.50 		& 3.221 				& 21.813   		& 1.955 				& 5.018 						&  68.0	\\
100				& 13.75 		& 2.713 				& 11.136	 		& 1.159 				& 3.546 						&  81.5	\\
100				& 15.00 		& 2.242 				& 5.911 			& 0.672 				& 2.455 						&  88.7	\\
100				& 16.25		& 1.882 				& 3.442 			& 0.398 				& 1.731 						&  92.6	\\
100				& 17.50		& 1.560 				& 2.128 			& 0.239 				& 1.221 						&  94.9	\\
\end{tabular}
\end{ruledtabular}
\label{table:TRIM_H2}

\end{table*}

An important conclusion is that to avoid large differences in detection efficiency, it is best to avoid using SB detectors with thick gold layers~\cite{Byrne_2022}, such as 100\,\microgcm\ SB detectors. If one does need to operate with a thick gold for the reason of studying a systematic effect, an 80\,\microgcm\ SB detector is preferable and should be run at an acceleration voltage of 30\,kV or larger.

This result provides an indication of how the neutron lifetime would be affected by the presence of \Htwop. For the majority of the apparatus configurations, there is little change to the neutron lifetime when compared to the total uncertainties of previous measurements of this type. As an example, consider a case where 5\,\% of the trapped events are \Htwop\ ions for a 60\,\microgcm\ SB detector at 30\,kV. In this scenario, the detection efficiency difference is 1.1\,\%.  This would equate to a change in the measured neutron lifetime of about 0.5\,s. Assuming a detection efficiency difference of 1.5\,\%, one would need to have 13\,\% of all trapped events be \Htwop\ in order to have a 1\,s change in the measured neutron lifetime. The actual amount of \Htwop\ seen in these data varies from 0\,\% to 5\,\%. 

\section{\label{sec:BL1Result}Estimation for the BL1 Lifetime Measurement}

The analysis to this point has come from data from the BL2 collaboration that have not been used to report a neutron lifetime result. It has demonstrated that it is possible to see the influence of \Htwo\ in both the timing and energy spectra of proton counting data that one would acquire in a cold-beam neutron lifetime experiment.  The natural question to ask at this point is whether or not charge exchange on \Htwo\ could have had an influence on the previously reported beam neutron lifetime~\cite{Dewey_2003,Yue2013}, referred to as BL1. While it is not possible to perform all of the same systematics tests on the BL1 data, one can perform some of the same analyses to search for evidence of \Htwo. Specifically, one can compare the shape of the timing spectrum with a simulation to look for excess events at late times, and one can examine the distribution of events in the 2-D timing versus energy plots.

\begin{figure}
  \centering
  \includegraphics[width=.45\textwidth]{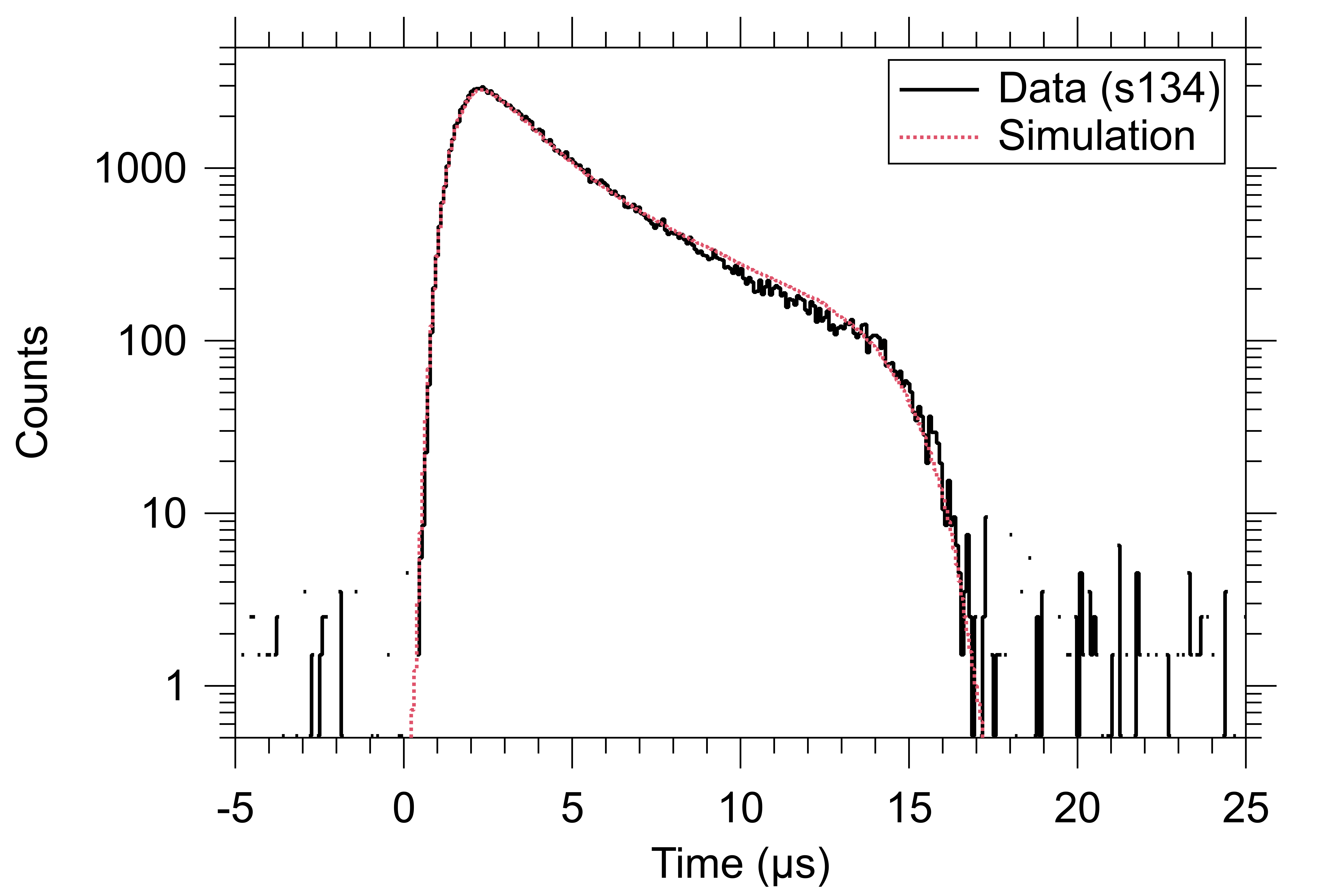}
  \caption{Proton arrival time spectra for data from the BL1 experiment and a GEANT4 simulation with no admixture of \Htwo. Background has been subtracted from the trace with data.}
  \label{fig:BL1timing}.      
\end{figure}

Figure~\ref{fig:BL1timing} shows a comparison of a typical timing spectrum from BL1 along with a GEANT4 simulation of the proton arrival time. The conditions were a trap length of 10 electrodes, a ramp voltage of 20\,V, a trapping time of 10\,ms, and an acceleration potential of 30\,kV. The trap length of 10 electrodes was chosen because the longer trap length increases the likelihood of observing \Htwop\ ions in the spectrum. Similarly, the 20 V ramp voltage (the lowest used in BL1), also increases the likelihood of observing \Htwop\ ions in the spectrum. The simulation assumes that there is no contribution from charge exchange on \Htwo, i.e., only protons are used in the simulation. This plot should be compared with Fig.~\ref{fig:windows_open_closed_timing_spectrum}, where there is a small contribution of \Htwo\, and yet it produces a clear signal in the timing spectrum. GEANT4 simulations were made for the longest trap lengths from all thirteen series from the BL1 data and similar plots to that shown in Fig.~\ref{fig:BL1timing} were made. The comparison of data with simulation does not show any indication of excess events at late times consistent with charge exchange with \Htwo\ molecules. We note that contrary to published assertions~\cite{Serebrov_2021}, all \Htwop\ ions would arrive well within the proton acceptance window~\cite{Nico_2005} and would be counted with high efficiency, as discussed in Section~\ref{sec:H2Measurement}.

Evidence for \Htwop\ in the proton counting appears more clearly in the 2-D timing versus energy plots, as seen in Fig.\ref{fig:RGA_data}. If it were in the BL1 data, one would expect it to appear more clearly in such plots as well. To investigate the potential influence of \Htwop, BL1 data were simulated using GEANT4 for the timing spectra and SRIM for the energy spectra. For the energy spectra, a SRIM simulation was run using incident proton energies ranging from 27.5\,keV to 32.5\,keV, corresponding to the experimental conditions. The protons were incident on a thin layer of gold or silicon to match the type of detector, SB or PIPS, respectively, that was used to acquire the data. The thickness of the gold layer came from the manufacturer's specification for the SB detector. There was no comparable specification for the PIPS detectors, and consequently, the silicon layer was varied until the energy loss of the simulation matched the data. SRIM matched the proton energy loss well but produced a narrower proton peak compared to the data. To account for this difference, approximately 2.9\,keV of Gaussian noise was added to each simulated spectrum to match the width of the data. This process was similar to that performed to simulate the \Htwop\ energy spectra as discussed in Section~\ref{sec:H2Measurement}.

To generate 2-D spectra, the energy and timing spectra used realistic noise and background as determined from the data. To generate 2-D spectra without any admixture of \Htwop, the simulated timing and energy spectra were used as probability density functions, and events were randomly generated with the same statistics as the data. A probability density function was also created using the off-peak section of the spectrum. This allowed for the appropriate background to be added to the simulated 2-D spectra. \Htwop\ spectra generated using GEANT4 could then be added to show how the 2-D spectra would appear with varying admixtures of \Htwop. With the use of the simulated energy and timing spectra and the use of the empirical background from the data, 2-D spectra were generated that mirror the data extremely well.

\begin{figure}
  \centering
  \includegraphics[width=.45\textwidth]{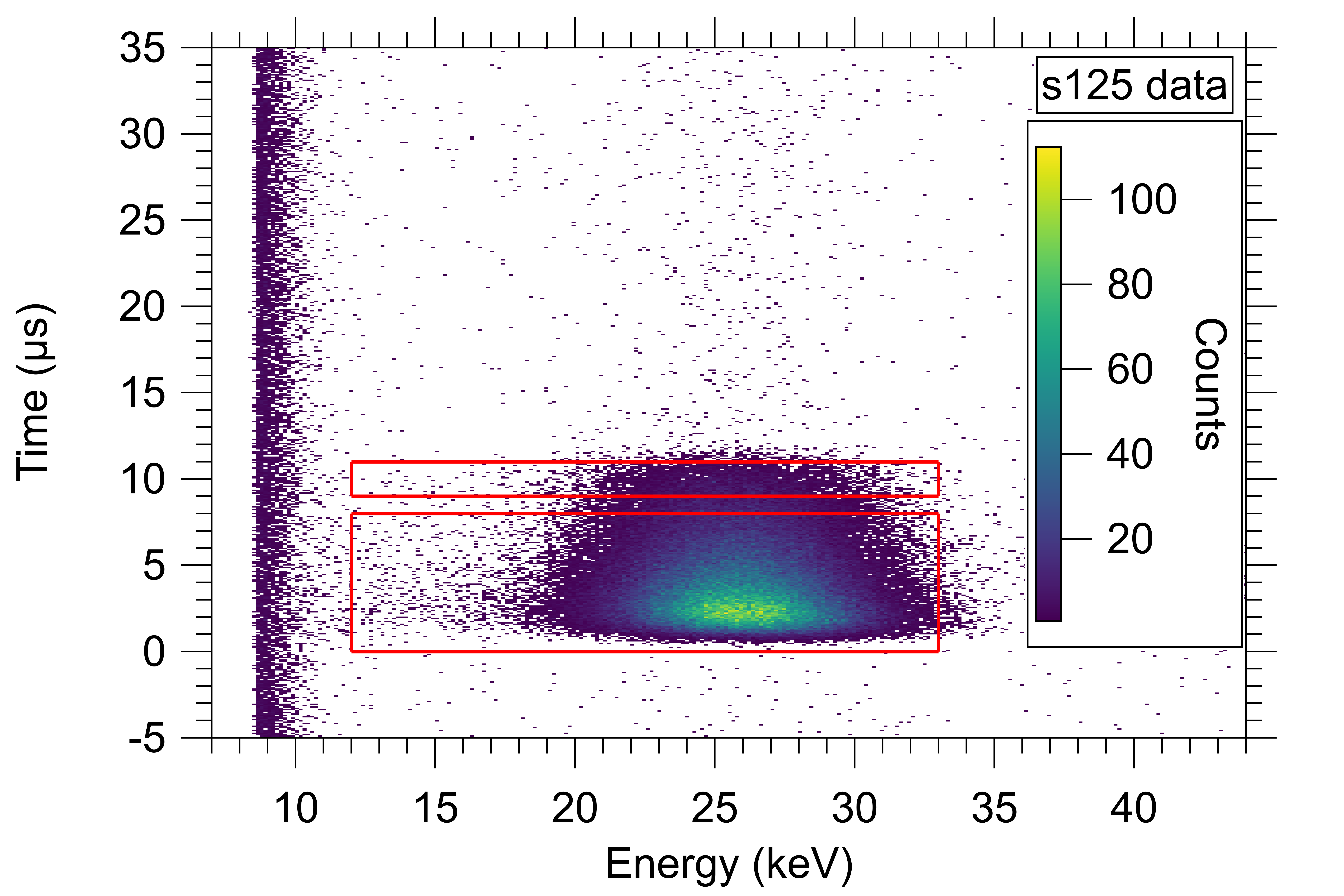}
  \includegraphics[width=.45\textwidth]{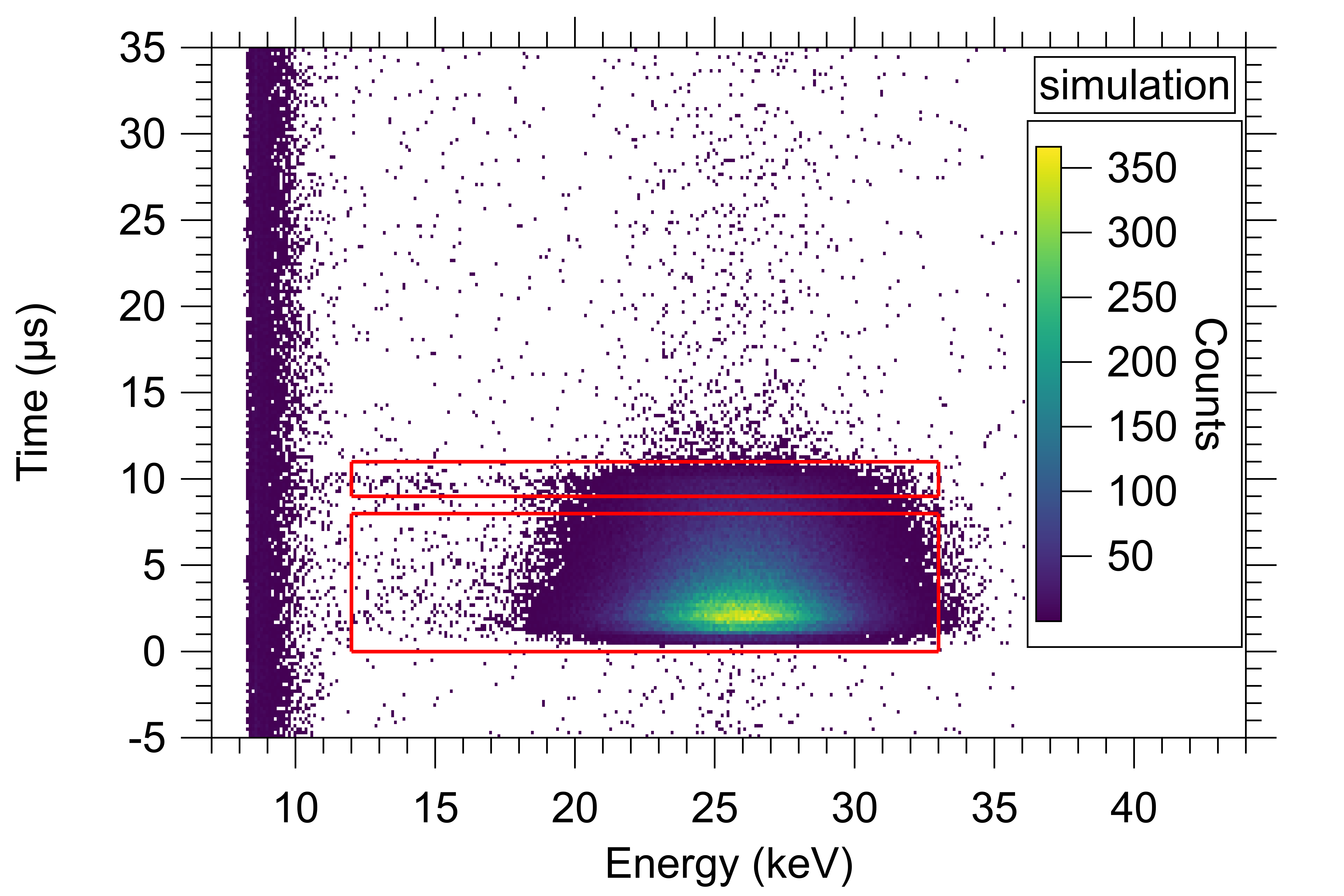}
  \caption{Top: timing versus energy histogram for data from one series of the BL1 data set.  Bottom: simulation of the same data set with a 1\,\% admixture of \Htwop\ ions. The red boxes on both plots are the two regions that were used to form ratios of counts.}
  \label{fig:BL1vGeant}
\end{figure}

To estimate the possible contribution of \Htwop, we compared the simulated 2-D spectra with given amounts of \Htwop\ to the BL1 data. First, with the simulated data the expected arrival time and deposited energy of the \Htwop\ ions were determined, as seen in Fig.~\ref{fig:BL1vGeant}. With that data, a ratio was made between the number of total events in this \Htwop\ region (the upper red box) to the number of protons that are in a cleaner region of the 2-D spectrum (the lower red box), i.e., a region less accessible to \Htwop\ ions. Second, this same timing cut was applied to the BL1 data, and the ratio was calculated between the expected \Htwop\ region and the proton-only region. This process was repeated with varying admixtures of \Htwop, and thus, a comparison could be made to determine which admixture best matched the BL1 data. This comparison of the simulated 2-D spectra to the BL1 data was performed for the longest trap length from each of the thirteen series used in the BL1 result. 

Using this method, there was no evidence for \Htwop\ in any of the BL1 data, but the detection limitation to this analysis is approximately 1\,\%. Using the same method as described in Section~\ref{subsec:DetEff}, the detection efficiency for protons and \Htwop\ was calculated for each apparatus configuration used in BL1 and tabulated in Table~\ref{table:TRIM_BL1}. One can estimate the effect on the measured neutron lifetime with the aforementioned analysis limitations as the upper bound for the possible \Htwop\ existence. Applying these upper limits in a worse-case scenario with the BL1 data set produces a shift in the lifetime of $-0.3$\,s. We note that even without this ratio analysis the amount of \Htwop\ required to shift the neutron lifetime by 1.0\,s would require an \Htwo\ contamination of 3.5\,\% or more and would be apparent by eye in the timing or 2-D spectra.

\begin{table*}[ht]
\caption{Table of proton and \Htwo\ efficiencies for the BL1 data. For both the experiment and the simulation, voltage is the proton acceleration voltage, $\rho_{\rm{Au}}$ is the areal density of the SB gold layer (a value of 0 implies a PIPS detector), and E$_{\rm{thr}}$  is the energy detection threshold. To the right of the vertical line are the results for the efficiencies, $\epsilon_{\rm p}$ and $\epsilon_{\rm H_2}$.}
\begin{ruledtabular}
\begin{tabular}{c c c c | c c c} 
 BL1 Series 	& Voltage\,(kV)	& $\rho_{\rm{Au}}$ \,($\mu$g/cm$^2$)	& E$_{\rm{thr}}$\,(keV)		&  \(\epsilon_{\rm p}\)\,(\%)	&  \(\epsilon_{\rm H_2}\)\,(\%)	& Difference\,(\%)  \\ [0.5ex]
 \hline
121			& 27.5		& 20				& 8.43				& 99.3				& 99.8				& -0.4 \\
125			& 27.5		& 20				& 8.43				& 99.3				& 99.8				& -0.4 \\
130			& 30.0		& 20				& 9.91				& 99.4				& 99.7				& -0.3 \\
134			& 30.0		& 20				& 9.91				& 99.4				& 99.7				& -0.3 \\
140			& 32.5		& 20				& 10.31				& 99.5				& 99.9				& -0.4 \\
142			& 32.5		& 20				& 10.31				& 99.5				& 99.9				& -0.4 \\
143			& 32.5		& 20				& 10.31				& 99.5				& 99.9				& -0.4 \\
149			& 27.5		& 60				& 10.66				& 96.7				& 85.3				& 11.4 \\
151			& 32.5		& 60				& 10.66				& 97.8				& 95.3				& 2.5 \\
154			& 30.0		& 0				& 11.22				& 99.6				& 93.6				& 6.0 \\
155			& 32.5		& 0				& 11.93				& 99.7				& 95.5				& 4.1 \\
166			& 27.5		& 40				& 9.89				& 98.3				& 96.3				& 2.0 \\
170			& 27.5		& 0				& 10.40				& 99.4				& 91.0				& 8.4 \\
\end{tabular}
\end{ruledtabular}
\label{table:TRIM_BL1}
\end{table*}

Given that the apparatus used in the studies for this paper is largely identical to that used in the BL1 experiment, one should ask why there are no similar instances of \Htwo\ in the BL1 data. There are two notable differences with the apparatus. The vacuum of the bore in BL1 was isolated from the beamline. For this work with the BL2 apparatus, the bore and the beamline shared vacuum. The beamline pressure was always higher than the bore because it was at room temperature, did not use ultrahigh vacuum hardware, and was not routinely baked at high temperatures. Isolating the two regions in BL1 eliminated the gas load that comes from the large volume of the beamline.  Downstream of the bore in BL1, a thin silicon wafer was placed in the beam path near the trap. Although it did not completely isolate the bore from the neutron counting region that was at higher pressure, it reduced the direct line-of-sight to the trap. These vacuum differences contributed to the decision to install the thin silicon windows in the BL2 apparatus, as discussed in Section~\ref{subsec:apparatus}.

Although we cannot entirely rule out the presence of trapped \Htwop\ in the BL1 data, it is not visible in any of the timing or 2-D spectra, such as those shown in Figs.~\ref{fig:BL1timing} and \ref{fig:BL1vGeant}. Furthermore, the efficiency for detecting \Htwop\ is high, and it would have been included along with the protons in the way the BL1 data were analyzed. Based on the quantitative estimates discussed in Section~\ref{subsec:DetEff}, we can state with confidence that charge exchange between trapped protons and molecular hydrogen would have produced a negligible (much less than 0.5\,s) effect on the reported neutron lifetime.

\section{\label{sec:Conclusion}Summary}

As part of studies of systematic effects that could affect the measured value of the neutron lifetime, we investigated the possibility of interactions with residual \Htwo\ gas causing a loss of protons. This is a concern for methods using beam techniques that trap and count the decay protons. Charge exchange with \Htwo\ in the BL2 beam-based neutron lifetime experiment was studied. It was demonstrated that, under certain experimental conditions, charge exchange can occur and is clearly observable in the data.  At the levels of precision that are of interest to the BL2 experiment, it is possible to significantly reduce the influence of \Htwo\ by maintaining a good vacuum and isolating the trap and superconducting magnet bore from the two regions of higher pressure in the apparatus. While it is always preferable to minimize the partial pressure of \Htwo\ in the trap, simulations show that the influence of comparatively large amounts of \Htwop\ contamination is mitigated because of its high detection probability.

The possible influence of \Htwo\ in the BL1 result of 2003 was also studied. Although there is no evidence of \Htwop\ in any of the data, the detection limit of this analysis technique was about 1\,\%. This limitation was due to higher noise and poorer energy resolution in the BL1 data as compared to the BL2 data. Regardless, analysis using a worst-case scenario for the amount of \Htwop\ contamination produces a shift in the measured neutron lifetime that is less than 0.5\,s.  We conclude that, while this effect is important to understand in order to fully characterize potential systematic effects, it is unlikely to be the cause of the current disagreement in neutron lifetime results. We do note that charge exchange producing \Htwop\ or other ions may be of particular concern for beam experiments aiming for higher precision, such as BL3~\cite{BL3_2021}. Care must be taken to understand the constituents of the residual gas and their interactions with the decay protons, and analysis techniques should be developed to identify the existence of ions other than protons in the data.

\begin{acknowledgments}

This work was supported by the National Institute of Standards and Technology (NIST), U.S. Department of Commerce; the US Department of Energy, Office of Nuclear Physics, under Interagency Agreement 89243023SSC000103 and grant DE-FG02-03ER41258; and the National Science Foundation grants PHY-2012395, PHY-2131864, and PHY-2309938, and the Cross-Disciplinary Science Institute at Gettysburg College. We acknowledge support from the NIST Center for Neutron Research, U.S. Department of Commerce, for providing the neutron facilities used in this work.

Identification of a product herein is for documentation purposes only, and does not imply recommendation or endorsement by NIST, nor does it imply that this product is necessarily the best available for the purpose.

\end{acknowledgments}

\nocite{data}

\bibliography{H2_v15}

@string{nima = "Nucl. Instrum. Meth. A"}

@phdthesis{Williams1989,
 	school = {University of Sussex},
	author = {Andreas P. Williams},
	title = {The Determination of the Neutron Lifetime by Trapping Decay Protons},
	year = {1989}
}

@article{Broussard2022,
	title = {Experimental Search for Neutron to Mirror Neutron Oscillations as an Explanation of the Neutron Lifetime Anomaly},
	author = {Broussard, L. J. and Barrow, J. L. and DeBeer-Schmitt, L. and Dennis, T. and Fitzsimmons, M. R. and Frost, M. J. and Gilbert, C. E. and Gonzalez, F. M. and Heilbronn, L. 	and Iverson, E. B. and Johnston, A. and Kamyshkov, Y. and Kline, M. and Lewiz, P. and Matteson, C. and Ternullo, J. and Varriano, L. and Vavra, S.},
	journal = {Phys. Rev. Lett.},
	volume = {128},
	issue = {21},
	pages = {212503},
	numpages = {7},
	year = {2022},
	month = {May},
	publisher = {American Physical Society},
	doi = {10.1103/PhysRevLett.128.212503},
	url = {https://link.aps.org/doi/10.1103/PhysRevLett.128.212503}
}

@article{Baym2018,
	title = {Testing Dark Decays of Baryons in Neutron Stars},
	author = {Baym, Gordon and Beck, D. H. and Geltenbort, Peter and Shelton, Jessie},
	journal = {Phys. Rev. Lett.},
	volume = {121},
	issue = {6},
	pages = {061801},
	numpages = {5},
	year = {2018},
	month = {Aug},
	publisher = {American Physical Society},
	doi = {10.1103/PhysRevLett.121.061801},
	url = {https://link.aps.org/doi/10.1103/PhysRevLett.121.061801}
}

@article{McKeen2018,
	title = {Neutron Stars Exclude Light Dark Baryons},
	author = {McKeen, David and Nelson, Ann E. and Reddy, Sanjay and Zhou, Dake},
	journal = {Phys. Rev. Lett.},
	volume = {121},
	issue = {6},
	pages = {061802},
	numpages = {5},
	year = {2018},
	month = {Aug},
	publisher = {American Physical Society},
	doi = {10.1103/PhysRevLett.121.061802},
	url = {https://link.aps.org/doi/10.1103/PhysRevLett.121.061802}
}

@article{Cline2018,
	title = {Dark decay of the neutron},
	author = {J.~M.~Cline and J.~M.~Cornell},
	journal = {J. High Energ. Phys.},
	volume = {2018},
	pages = {81},
	year = {2018}
}

@article{Motta2018,
	doi = {10.1088/1361-6471/aab689},
	url = {https://dx.doi.org/10.1088/1361-6471/aab689},
	year = {2018},
	month = {apr},
	publisher = {IOP Publishing},
	volume = {45},
	number = {5},
	pages = {05LT01},
	author = {Motta, T F and Guichon, P A M and Thomas, A W},
	title = {Implications of neutron star properties for the existence of light dark matter},
	journal = {Journal of Physics G: Nuclear and Particle Physics}
}

@article{Joubioux2024,
	title = {Search for a Neutron Dark Decay in $^{6}\mathrm{He}$},
	author = {Le Joubioux, M. and Savajols, H. and Mittig, W. and Fl\'echard, X. and Hayen, L. and Penionzhkevich, Yu. E. and Ackermann, D. and Borcea, C. and Caceres, L. and 		Delahaye, P. and Didierjean, F. and Franchoo, S. and Grillet, A. and Jacquot, B. and Lebois, M. and Ledoux, X. and Lecesne, N. and Li\'enard, E. and Lukyanov, S. and Naviliat-Cuncic, 	O. and Piot, J. and Singh, A. and Smirnov, V. and Stodel, C. and Testov, D. and Thisse, D. and Thomas, J. C. and Verney, D.},
	journal = {Phys. Rev. Lett.},
	volume = {132},
	issue = {13},
	pages = {132501},
	numpages = {6},
	year = {2024},
	month = {Mar},
	publisher = {American Physical Society},
	doi = {10.1103/PhysRevLett.132.132501},
	url = {https://link.aps.org/doi/10.1103/PhysRevLett.132.132501}
}

@article{Hassan2021,
    author = {{Hassan}, Md and {Floyd}, Nicholas and {Tang}, Zhaowen and {UCNProBe Team}},
    title = "{The current status of the UCNProBe experiment.}",
    journal = {Bull. Am. Phys. Soc.},
    volume = {66},
    year = {2021}
}

@article{Sun2018,
	title = {Search for dark matter decay of the free neutron from the {UCNA} experiment: $n\ensuremath{\rightarrow}\ensuremath{\chi}+{e}^{+}{e}^{\ensuremath{-}}$},
	author = {Sun, X. and Adamek, E. and Allgeier, B. and Blatnik, M. and Bowles, T. J. and Broussard, L. J. and Brown, M. A.-P. and Carr, R. and Clayton, S. and Cude-Woods, C. and 	Currie, S. and Dees, E. B. and Ding, X. and Filippone, B. W. and Garc\'{\i}a, A. and Geltenbort, P. and Hasan, S. and Hickerson, K. P. and Hoagland, J. and Hong, R. and Hogan, G. E. 	and Holley, A. T. and Ito, T. M. and Knecht, A. and Liu, C.-Y. and Liu, J. and Makela, M. and Mammei, R. and Martin, J. W. and Melconian, D. and Mendenhall, M. P. and Moore, S. D. 	and Morris, C. L. and Nepal, S. and Nouri, N. and Pattie, R. W. and P\'erez Galv\'an, A. and Phillips, D. G. and Picker, R. and Pitt, M. L. and Plaster, B. and Ramsey, J. C. and Rios, R. 	and Salvat, D. J. and Saunders, A. and Sondheim, W. and Sjue, S. and Slutsky, S. and Swank, C. and Swift, G. and Tatar, E. and Vogelaar, R. B. and VornDick, B. and Wang, Z. and 	Wei, W. and Wexler, J. and Womack, T. and Wrede, C. and Young, A. R. and Zeck, B. A.},
	collaboration = {UCNA Collaboration},
	journal = {Phys. Rev. C},
	volume = {97},
	issue = {5},
	pages = {052501},
	numpages = {5},
	year = {2018},
	month = {May},
	publisher = {American Physical Society},
	doi = {10.1103/PhysRevC.97.052501},
	url = {https://link.aps.org/doi/10.1103/PhysRevC.97.052501}
}

@article{Tang2018,
	title = {Search for the Neutron Decay $n\ensuremath{\rightarrow}X+\ensuremath{\gamma}$, Where $X$ is a Dark Matter Particle},
	author = {Tang, Z. and Blatnik, M. and Broussard, L. J. and Choi, J. H. and Clayton, S. M. and Cude-Woods, C. and Currie, S. and Fellers, D. E. and Fries, E. M. and Geltenbort, P. and 	Gonzalez, F. and Hickerson, K. P. and Ito, T. M. and Liu, C.-Y. and MacDonald, S. W. T. and Makela, M. and Morris, C. L. and O'Shaughnessy, C. M. and Pattie, R. W. and Plaster, B. 	and Salvat, D. J. and Saunders, A. and Wang, Z. and Young, A. R. and Zeck, B. A.},
	journal = {Phys. Rev. Lett.},
	volume = {121},
	issue = {2},
	pages = {022505},
	numpages = {5},
	year = {2018},
	month = {Jul},
	publisher = {American Physical Society},
	doi = {10.1103/PhysRevLett.121.022505},
	url = {https://link.aps.org/doi/10.1103/PhysRevLett.121.022505}
}

@misc{Blatnik2024,
	title={An experimental search for an explanation of the difference between beam and bottle neutron lifetime measurements}, 
	author={M. F. Blatnik and L. S. Blokland and N. Callahan and J. H. Choi and S. Clayton and C. B Cude-Woods and B. W. Filippone and W. R. Fox and E. Fries and P. Geltenbort and F. 	M. Gonzalez and L. Hayen and K. P. Hickerson and A. T. Holley and T. M. Ito and A. Komives and S Lin and Chen-Yu Liu and M. F. Makela and C. L. Morris and R. Musedinovic and C. 	M. O'Shaughnessy and R. W. Pattie Jr. and J. C. Ramsey and D. J. Salvat and A. Saunders and S. J. Seestrom and E. I. Sharapov and M. Singh and Z. Tang and W. F. Uhrich and J. 	Vanderwerp and P. Walstrom and Z. Wang and A. R. Young},
	year={2024},
	eprint={2406.10378},
	archivePrefix={arXiv},
	primaryClass={nucl-ex},
	url={https://arxiv.org/abs/2406.10378}, 
}

@article{Dubbers2019,
	title = {Exotic decay channels are not the cause of the neutron lifetime anomaly},
	journal = {Physics Letters B},
	volume = {791},
	pages = {6-10},
	year = {2019},
	issn = {0370-2693},
	doi = {https://doi.org/10.1016/j.physletb.2019.02.013},
	url = {https://www.sciencedirect.com/science/article/pii/S0370269319301066},
	author={Dirk Dubbers and Heiko Saul and Bastian M{\"a}rkisch and Torsten Soldner and Hartmut Abele}
}

@article{Klopf2019,
	title = {Constraints on the Dark Matter Interpretation $n\ensuremath{\rightarrow}\ensuremath{\chi}+{e}^{+}{e}^{\ensuremath{-}}$ of the Neutron Decay Anomaly with the {PERKEO II} 		Experiment},
	author = {Klopf, M. and Jericha, E. and M\"arkisch, B. and Saul, H. and Soldner, T. and Abele, H.},
	journal = {Phys. Rev. Lett.},
	volume = {122},
	issue = {22},
	pages = {222503},
	numpages = {4},
	year = {2019},
	month = {Jun},
	publisher = {American Physical Society},
	doi = {10.1103/PhysRevLett.122.222503},
	url = {https://link.aps.org/doi/10.1103/PhysRevLett.122.222503}
}

@article{Koch2024,
	title = {Exciting hint toward the solution of the neutron lifetime puzzle},
	author = {Koch, Benjamin and Hummel, Felix},
	journal = {Phys. Rev. D},
	volume = {110},
	issue = {7},
	pages = {073004},
	numpages = {11},
	year = {2024},
	month = {Oct},
	publisher = {American Physical Society},
	doi = {10.1103/PhysRevD.110.073004},
	url = {https://link.aps.org/doi/10.1103/PhysRevD.110.073004}
}

@article{Rajendran2021,
	title = {Composite solution to the neutron lifetime anomaly},
	author = {Rajendran, Surjeet and Ramani, Harikrishnan},
	journal = {Phys. Rev. D},
	volume = {103},
	issue = {3},
	pages = {035014},
	numpages = {8},
	year = {2021},
	month = {Feb},
	publisher = {American Physical Society},
	doi = {10.1103/PhysRevD.103.035014},
	url = {https://link.aps.org/doi/10.1103/PhysRevD.103.035014}
}

@article{Fornal2018,
	title = {Dark Matter Interpretation of the Neutron Decay Anomaly},
	author = {Fornal, Bartosz and Grinstein, Benjam\'{\i}n},
	journal = {Phys. Rev. Lett.},
	volume = {120},
	issue = {19},
	pages = {191801},
	numpages = {6},
	year = {2018},
	month = {May},
	publisher = {American Physical Society},
	doi = {10.1103/PhysRevLett.120.191801},
	url = {https://link.aps.org/doi/10.1103/PhysRevLett.120.191801}
}

@misc{Fuwa2024,
	title={Improved measurements of neutron lifetime with cold neutron beam at {J-PARC}}, 
	author={Y. Fuwa and T. Hasegawa and K. Hirota and T. Hoshino and R. Hosokawa and G. Ichikawa and S. Ieki and T. Ino and Y. Iwashita and M. Kitaguchi and R. Kitahara and S. 		Makise and K. Mishima and T. Mogi and N. Nagakura and H. Oide and H. Okabe and H. Otono and Y. Seki and D. Sekiba and T. Shima and H. E. Shimizu and H. M. Shimizu and N. 	Sumi and H. Sumino and M. Tanida and H. Uehara and T. Yamada and S. Yamashita and K. Yano and T. Yoshioka},
	year={2024},
	eprint={2412.19519},
	archivePrefix={arXiv},
	primaryClass={nucl-ex},
	url={https://arxiv.org/abs/2412.19519}, 
}

@article{ Materne2009,
	Author = {Materne, S. and Picker, R. and Altarev, I. and Angerer, H. and Franke, B. and Gutsmiedl, E. and Hartmann, F. J. and Mueller, A. R. and Paul, S.	 and Stoepler, R.},
	Title = {PENeLOPE-on the way towards a new neutron lifetime experiment with magnetic storage of ultra-cold neutrons and proton extraction},
	Journal = nima,
	Year = {2009},
	Volume = {611},
	Number = {2-3},
	Pages = {176-180},
	Month = {DEC 1},
	Note = {5th International Workshop on Particle Physics with Slow Neutrons, Inst Laue Langevin, Grenoble, FRANCE, MAY 29-31, 2008},
	Organization = {Inst Natl Phys Nucl \& Phys Particules, CNRS; Univ Joseph Fourier; Excellence Cluster Origin \& Struct Universe},
	DOI = {10.1016/j.nima.2009.07.055},
	ISSN = {0168-9002},
	ResearcherID-Numbers = {Franke, Beatrice/B-9809-2017 Paul, Stephan/F-7596-2015},
	ORCID-Numbers = {Franke, Beatrice/0000-0001-8417-4603 Paul, Stephan/0000-0002-8813-0437 Picker, Rudiger/0000-0002-7668-0693},
	Unique-ID = {WOS:000273101500015},
}

@article{Yue2013,
	title = {Improved Determination of the Neutron Lifetime},
	author = {Yue, A. T. and Dewey, M. S. and Gilliam, D. M. and Greene, G. L. and Laptev, A. B. and Nico, J. S. and Snow, W. M. and Wietfeldt, F. E.},
  	journal = {Phys. Rev. Lett.},
  	volume = {111},
  	issue = {22},
  	pages = {222501},
  	numpages = {4},
  	year = {2013},
  	month = {Nov},
  	publisher = {American Physical Society},
  	doi = {10.1103/PhysRevLett.111.222501},
  	url = {https://link.aps.org/doi/10.1103/PhysRevLett.111.222501}
}

@article{Hoogerheide2019,
	author = {{Hoogerheide, Shannon F.} and {Caylor, Jimmy} and {Adamek, Evan R.} and {Anderson, Eamon S.} and {Biswas, Ripan} and {Chavali, Sai Meghasena} and {Crawford, Bret} and {DeAngelis, 		Christina} and {Dewey, Maynard S.} and {Fomin, Nadia} and {Gilliam, David M.} and {Grammer, Kyle B.} and {Greene, Geoffrey L.} and {Haun, Robert W.} and {Ivanov, Juliet A.} and {Li, Fangchen} and 		{Mulholland, Jonathan} and {Mumm, H. Pieter} and {Nico, Jeffrey S.} and {Snow, William M.} and {Valete, Daniel} and {Wietfeldt, Fred E.} and {Yue, Andrew T.}},
	title = {Progress on the {BL2} beam measurement of the neutron lifetime},
	DOI= "10.1051/epjconf/201921903002",
	url= "https://doi.org/10.1051/epjconf/201921903002",
	journal = {EPJ Web Conf.},
	year = 2019,
	volume = 219,
	pages = "03002",
}

@unpublished{BL3_2021,
	author = "{BL3 Collaboration}",
	title = {{BL3} Project Conceptual Design Report},
	note = {unpublished},
	institution = {Tulane University},
	year = {2021},
}

@software{MATLAB,
	year = {2022},
	author = {The MathWorks Inc.},
	title = {MATLAB version: 9.13.0 (R2022b)},
	publisher = {The MathWorks Inc.},
	address = {Natick, Massachusetts, United States},
	url = {https://www.mathworks.com}
}

@ARTICLE{Allison2006,
	author={Allison, J. and Amako, K. and Apostolakis, J. and Araujo, H. and Arce Dubois, P. and Asai, M. and Barrand, G. and Capra, R. and Chauvie, S. 		and Chytracek, R. and Cirrone, 	G.A.P. and Cooperman, G. and Cosmo, G. and Cuttone, G. and Daquino, G.G. and Donszelmann, M. and Dressel, M. and Folger, G. and Foppiano, F. and Generowicz, J. and 				Grichine, V. and Guatelli, S. and Gumplinger, P. and Heikkinen, A. and Hrivnacova, I. and 	Howard, A. and Incerti, S. and Ivanchenko, V. and Johnson, T. and Jones, F. and Koi, T. and 			Kokoulin, R. and Kossov, M. and Kurashige, H. and Lara, V. 	and Larsson, S. and Lei, F. and Link, O. and Longo, F. and Maire, M. and Mantero, A. and Mascialino, B. and McLaren, I. 			and Mendez Lorenzo, P. and Minamimoto, K. and Murakami, K. and Nieminen, P. and Pandola, L. and Parlati, S. and Peralta, L. and Perl, J. and Pfeiffer, A. and Pia, M.G. and Ribon, 			A. and Rodrigues, P. and Russo, G. and Sadilov, S. and Santin, G. and Sasaki, T. and Smith, D. and Starkov, N. and Tanaka, S. and Tcherniaev, E. and 		Tome, B. and Trindade, A. and 		Truscott, P. and Urban, L. and Verderi, M. and Walkden, A. and Wellisch, J.P. and Williams, D.C. and Wright, D. and 	Yoshida, H.},
  	journal={IEEE Transactions on Nuclear Science}, 
  	title={Geant4 developments and applications}, 
  	year={2006},
  	volume={53},
  	number={1},
  	pages={270-278}
}

@article{Agostinelli2003,
	title = {Geant4—a simulation toolkit},
	journal = {Nuclear Instruments and Methods in Physics Research Section A: Accelerators, Spectrometers, Detectors and Associated Equipment},
	volume = {506},
	number = {3},
	pages = {250-303},
	year = {2003},
	issn = {0168-9002},
	author = {S. Agostinelli and J. Allison and K. Amako and J. Apostolakis and H. Araujo and P. Arce and M. Asai and D. Axen and S. Banerjee and G. 			Barrand and F. Behner and L. Bellagamba and J. Boudreau and L. Broglia and A. Brunengo and H. Burkhardt and S. Chauvie and J. Chuma and R. 		Chytracek and G. Cooperman and G. Cosmo and P. Degtyarenko and A. Dell'Acqua and G. Depaola and D. Dietrich and R. Enami and A. Feliciello and 		C. Ferguson and H. Fesefeldt and G. Folger and F. Foppiano and A. Forti and S. Garelli and S. Giani and R. Giannitrapani and D. Gibin and J.J. {Gómez 		Cadenas} and I. González and G. {Gracia Abril} and G. Greeniaus and W. Greiner and V. Grichine and A. Grossheim and S. Guatelli and P. Gumplinger 		and R. Hamatsu and K. Hashimoto and H. Hasui and A. Heikkinen and A. Howard and V. Ivanchenko and A. Johnson and F.W. Jones and J. Kallenbach 		and N. Kanaya and M. Kawabata and Y. Kawabata and M. Kawaguti and S. Kelner and P. Kent and A. Kimura and T. Kodama and R. Kokoulin and M. 		Kossov and H. Kurashige and E. Lamanna and T. Lampén and V. Lara and V. Lefebure and F. Lei and M. Liendl and W. Lockman and F. Longo and S. 		Magni and M. Maire and E. Medernach and K. Minamimoto and P. {Mora de Freitas} and Y. Morita and K. Murakami and M. Nagamatu and R. Nartallo 		and P. Nieminen and T. Nishimura and K. Ohtsubo and M. Okamura and S. O'Neale and Y. Oohata and K. Paech and J. Perl and A. Pfeiffer and M.G. Pia 	and F. Ranjard and A. Rybin and S. Sadilov and E. {Di Salvo} and G. Santin and T. Sasaki and N. Savvas and Y. Sawada and S. Scherer and S. Sei and 		V. Sirotenko and D. Smith and N. Starkov and H. Stoecker and J. Sulkimo and M. Takahata and S. Tanaka and E. Tcherniaev and E. {Safai Tehrani} and 		M. Tropeano and P. Truscott and H. Uno and L. Urban and P. Urban and M. Verderi and A. Walkden and W. Wander and H. Weber and J.P. Wellisch and 		T. Wenaus and D.C. Williams and D. Wright and T. Yamada and H. Yoshida and D. Zschiesche}
}

@article{Allison2016,
	title = {Recent developments in Geant4},
	journal = {Nuclear Instruments and Methods in Physics Research Section A: Accelerators, Spectrometers, Detectors and Associated Equipment},
	volume = {835},
	pages = {186-225},
	year = {2016},
	issn = {0168-9002},
	author = {J. Allison and K. Amako and J. Apostolakis and P. Arce and M. Asai and T. Aso and E. Bagli and A. Bagulya and S. Banerjee and G. Barrand 		and B.R. Beck and A.G. Bogdanov and D. Brandt and J.M.C. Brown and H. Burkhardt and Ph. Canal and D. Cano-Ott and S. Chauvie and K. Cho and 		G.A.P. Cirrone and G. Cooperman and M.A. Cortés-Giraldo and G. Cosmo and G. Cuttone and G. Depaola and L. Desorgher and X. Dong and A. Dotti 		and V.D. Elvira and G. Folger and Z. Francis and A. Galoyan and L. Garnier and M. Gayer and K.L. Genser and V.M. Grichine and S. Guatelli and P. 		Guèye and P. Gumplinger and A.S. Howard and I. Hřivnáčová and S. Hwang and S. Incerti and A. Ivanchenko and V.N. Ivanchenko and F.W. Jones and 		S.Y. Jun and P. Kaitaniemi and N. Karakatsanis and M. Karamitros and M. Kelsey and A. Kimura and T. Koi and H. Kurashige and A. Lechner and S.B. 		Lee and F. Longo and M. Maire and D. Mancusi and A. Mantero and E. Mendoza and B. Morgan and K. Murakami and T. Nikitina and L. Pandola and P. 		Paprocki and J. Perl and I. Petrović and M.G. Pia and W. Pokorski and J.M. Quesada and M. Raine and M.A. Reis and A. Ribon and A. {Ristić Fira} and 		F. Romano and G. Russo and G. Santin and T. Sasaki and D. Sawkey and J.I. Shin and I.I. Strakovsky and A. Taborda and S. Tanaka and B. Tomé and T. 	Toshito and H.N. Tran and P.R. Truscott and L. Urban and V. Uzhinsky and J.M. Verbeke and M. Verderi and B.L. Wendt and H. Wenzel and D.H. Wright 		and D.M. Wright and T. Yamashita and J. Yarba and H. Yoshida}
}

@misc{FunPhysWhite2023,
	author = {Alarcon, Ricardo and others},
	year = {2023},
	month = {08},
	pages = {},
	title = {Fundamental Neutron Physics: a White Paper on Progress and Prospects in the {US}}
}

@article{Byrne1989,
	title = {Determination of the neutron lifetime by counting trapped protons},
	journal = {Nuclear Instruments and Methods in Physics Research Section A: Accelerators, Spectrometers, Detectors and Associated Equipment},
	volume = {284},
	number = {1},
	pages = {116-119},
	year = {1989},
	issn = {0168-9002},
	doi = {https://doi.org/10.1016/0168-9002(89)90261-1},
	url = {https://www.sciencedirect.com/science/article/pii/0168900289902611},
	author = {J. Byrne and P. G. Dawber and J. A. Spain and M. S. Dewey and D. M. Gilliam and G. L. Greene and G. P. Lamaze and A. P. Williams and J. Pauwels and R. Eykens and J. VanGestel and A. 		Lamberty and R. D. Scott}
}

@article{Lawrence2021,
	title = {Space-based measurements of neutron lifetime: Approaches to resolving the neutron lifetime anomaly},
	journal = {Nuclear Instruments and Methods in Physics Research Section A: Accelerators, Spectrometers, Detectors and Associated Equipment},
	volume = {988},
	pages = {164919},
	year = {2021},
	issn = {0168-9002},
	doi = {https://doi.org/10.1016/j.nima.2020.164919},
	url = {https://www.sciencedirect.com/science/article/pii/S0168900220313164},
	author = {David J. Lawrence and Jack T. Wilson and Patrick N. Peplowski}
}

@article{Hirota2020,
	author = {Hirota, K and Ichikawa, G and Ieki, S and Ino, T and Iwashita, Y and Kitaguchi, M and Kitahara, R and Koga, J and Mishima, K and Mogi, T and 	Morikawa, K and Morishita, A and Nagakura, N and Oide, H and Okabe, H and Otono, H and Seki, Y and Sekiba, D and Shima, T and Shimizu, H M and 		Sumi, N and Sumino, H and Tomita, T and Uehara, H and Yamada, T and Yamashita, S and Yano, K and Yokohashi, M and Yoshioka, T},
	title = "{Neutron lifetime measurement with pulsed cold neutrons}",
	journal = {Progress of Theoretical and Experimental Physics},
	volume = {2020},
	number = {12},
	pages = {123C02},
	year = {2020},
	month = {12},
	issn = {2050-3911},
	eprint = {https://academic.oup.com/ptep/article-pdf/2020/12/123C02/35931162/ptaa169.pdf}
}

@phdthesis{Caylor2022,
 	school = {The University of Tennessee, Knoxville},
	author = {Jimmy Caylor},
	title = {The Upgraded Measurement of the Neutron Lifetime Using the In-Beam Method.},
	year = {2022},
	url = {https://trace.tennessee.edu/utk_graddiss/7195}
}

@article{trapfilter,
	author = {V. T. Jordanov and G. F. Knoll},      
	title = {Digital synthesis of pulse shapes in real time for high resolution radiation spectroscopy},
	journal = {Nuclear Instruments and Methods in Physics Research Section A},           
	volume = {345},
	number = {2},pages = {337-345},
	year = {1994}    
}

@article{Yue_2018,
	doi = 		{10.1088/1681-7575/aac283},
	url = 			{https://doi.org/10.1088/1681-7575/aac283},
	year = 		2018,
	month = 		{jun},
	publisher = 	{{IOP} Publishing},
	volume = 		{55},
	number = 		{4},
	pages = 		{460--485},
	author = 		{A T Yue and E S Anderson and M S Dewey and D M Gilliam and G L Greene and A B Laptev and J S Nico and W M Snow},
	title = 		{Precision determination of absolute neutron flux},
	journal = 		{Metrologia},
}

@book{BarCohen2016,
  	title=			{Low Temperature Materials and Mechanisms},
  	author=		{Y. Bar-Cohen},
  	isbn=		{9781498700399},
  	lccn=		{2016000772},
  	url=			{https://books.google.com/books?id=8MHBDAAAQBAJ},
  	year=		{2016},
  	publisher=		{CRC Press},
    address = {Boca Raton}
}

@article{Serebrov_2021,
  	title = 		{Search for explanation of the neutron lifetime anomaly},
  	author = 		{Serebrov, A. P. and Chaikovskii, M. E. and Klyushnikov, G. N. and Zherebtsov, O. M. and Chechkin, A. V.},
  	journal = 		{Phys. Rev. D},
  	volume = 		{103},
  	issue = 		{7},
  	pages = 		{074010},
  	numpages = 	{17},
  	year = 		{2021},
  	month = 		{Apr},
  	publisher = 	{American Physical Society},
  	doi = 		{10.1103/PhysRevD.103.074010},
  	url = 			{https://link.aps.org/doi/10.1103/PhysRevD.103.074010}
}

@article{Byrne_2019,
	author = 		{J. Byrne and D. L. Worcester},
	title = 			"{The neutron lifetime anomaly and charge exchange collisions of trapped protons}",
	doi = 			"10.1088/1361-6471/ab256b",
	journal = 		"J. Phys. G",
	volume =		"46",
	number = 		"8",
	pages = 		"085001",
	year = 			"2019"
}

@article{Byrne_1996,
  	title		= 	{A revised value for the neutron lifetime measured using a {P}enning trap},
  	author	= 	{J. Byrne and P. G. Dawber and C. G. Habeck and S. J. Smidt and J. A. Spain and A. P. Williams},
  	journal	= 	{Europhys. Lett.},
  	year		= 	{1996},
  	number	= 	{3},
  	pages	= 	{187},
  	volume	= 	{33},
  	doi		= 	{10.1209/epl/i1996-00319-x}
}

@article{Gilbody_1957,
	author = 		{H. B. Gilbody and J. B. Hasted},
	title = 		"{Anomalies in the adiabatic interpretation of charge-transfer collisions}",
	doi = 		"https://doi.org/10.1098/rspa.1957.0004",
	journal = 		"Proceedings of the Royal Society A",
	volume = 		"238",
	number = 		"1214",
	year = 		"1957"
}

@article{AtomicData_1978,
	title = 		{Cross sections for charge transfer of hydrogen beams in gases and vapors in the energy range 10 e{V} - 10 ke{V}},
	journal = 		{Atomic Data and Nuclear Data Tables},
	volume = 		{22},
	number = 		{6},
	pages = 		{491-525},
	year = 		{1978},
	issn = 		{0092-640X},
	doi = 		{https://doi.org/10.1016/0092-640X(78)90021-9},
	url = 			{https://www.sciencedirect.com/science/article/pii/0092640X78900219},
	author = 		{H. Tawara}
}

@article{Allison_1958,
  	title = 		{Experimental {R}esults on {C}harge-{C}hanging {C}ollisions of {H}ydrogen and {H}elium {A}toms and {I}ons at {K}inetic {E}nergies above 0.2 ke{V}},
  	author = 		{Allison, Samuel K.},
  	journal = 		{Rev. Mod. Phys.},
  	volume = 		{30},
  	issue = 		{4},
  	pages = 		{1137--1168},
  	numpages = 	{0},
  	year = 		{1958},
  	month = 		{Oct},
  	publisher = 	{American Physical Society},
  	doi = 		{10.1103/RevModPhys.30.1137},
  	url = 			{https://link.aps.org/doi/10.1103/RevModPhys.30.1137}
}

@article{Byrne_1990,
  	title = 		{Measurement of the neutron lifetime by counting trapped protons},
  	author = 		{Byrne, J. and Dawber, P. G. and Spain, J. A. and Williams, A. P. and Dewey, M. S. and Gilliam, D. M. and Greene, G. L. and Lamaze, G. 		P. and Scott, R. D. and Pauwels, J. and Eykens, R. and Lamberty, A.},
  	journal = 		{Phys. Rev. Lett.},
  	volume = 		{65},
  	issue = 		{3},
  	pages = 		{289--292},
  	numpages = 	{0},
  	year = 		{1990},
  	month = 		{Jul},
  	publisher = 	{American Physical Society},
  	doi = 		{10.1103/PhysRevLett.65.289},
  	url = 			{http://link.aps.org/doi/10.1103/PhysRevLett.65.289}
}

@article{Dewey_2003,
  	title = 		{Measurement of the Neutron Lifetime Using a Proton Trap},
  	author = 		{Dewey, M. S. and Gilliam, D. M. and Nico, J. S. and Wietfeldt, F. E. and Fei, X. and Snow, W. M. and Greene, G. L. and Pauwels, J. and Eykens, R. and Lamberty, A. and Van Gestel, 					J.},
  	journal = 		{Phys. Rev. Lett.},
  	volume = 		{91},
  	issue = 		{15},
  	pages = 		{152302},
  	numpages = 	{4},
  	year = 		{2003},
  	month = 		{Oct},
  	publisher = 	{American Physical Society},
  	doi = 		{10.1103/PhysRevLett.91.152302},
  	url = 			{http://link.aps.org/doi/10.1103/PhysRevLett.91.152302}
}

@article{Nico_2005,
	author = 		"J. S. Nico and M. S. Dewey and D. M. Gilliam and F.E. Wietfeldt and X. Fei and W. M. Snow and G. L. Greene and J. Pauwels and R. Eykens and A. Lamberty and J. {Van Gestel} and R. D. Scott",
	title = 		"{Measurement of the neutron lifetime by counting trapped protons in a cold neutron beam}",
	eprint =		"nucl-ex/0411041",
	archivePrefix =	"arXiv",
	doi =			"10.1103/PhysRevC.71.055502",
	journal =		"Phys. Rev. C",
	volume = 		"71",
	pages =		"055502",
	year =		"2005"
}

@article{Wietfeldt_2011,
	author = 		{Fred E. Wietfeldt and Geoffrey L. Greene},
	title = 		"{Colloquium: The neutron lifetime}",
	doi = 		"10.1103/RevModPhys.83.1173",
	journal = 		"Rev. Mod. Phys.",
	volume = 		"83",
	number = 		"4",
	pages = 		"1173--1192",
	year = 		"2011"
}

@article{Stedeford_1955,
	author = 		{J. B. H. Stedeford and J. B. Hasted},
	title = 		{Further investigations of charge exchange and electron detachment -  {I}. {I}on energies 3 to 40 ke{V} -  {II}. {I}on energies 100 to 4000 		e{V}},
	doi = 		{https://doi.org/10.1098/rspa.1955.0024},
	journal = 		{Proceedings of the Royal Society A},
	volume = 		"227",
	number = 		"1171",
	year = 		"1955"
}

@article{Phelps_1990,
	author = 		{A. V. Phelps},
	title = 		{Cross Sections and Swarm Coefficients for {H$^+$}, {H$_2^+$}, {H$_3^+$}, {H}, {H$_2$}, and {H$^-$} in {H$_2$} for Energies from 0.1 e{V} to 10 ke{V}},
	journal = 		{Journal of Physical and Chemical Reference Data},
	volume = 		{19},
	pages = 		{653-675},
	year = 		{1990},
	URL = 		{https://doi.org/10.1063/1.555858}
}

@article{Phelps_1992,
	author = 		{A. V. Phelps},
	title = 		{Collisions of {H$^+$}, {H$_2^+$}, {H$_3^+$}, {ArH$^+$}, {H$^-$}, {H}, and {H$_2$} with {Ar} and of {Ar$^+$} and {ArH$^+$} with {H$_2$} for {E}nergies from 0.1 		e{V} to 10 ke{V}},
	journal = 		{Journal of Physical and Chemical Reference Data},
	volume = 		{21},
	number = 		{4},
	pages = 		{883-897},
	year = 		{1992},
	doi = 		{10.1063/1.555917},
	URL = 		{https://doi.org/10.1063/1.555917}
}

@article{Stebbings_1964,
	author = {R. F. Stebbings and A. C. H. Smith and H. Ehrhardt},
	title = {Charge transfer between oxygen atoms and {O$^+$} and {H$^+$} ions},
	journal = {Journal of Geophysical Research (1896-1977)},
	volume = {69},
	number = {11},
	pages = {2349-2355},
	doi = {https://doi.org/10.1029/JZ069i011p02349},
	url = {https://agupubs.onlinelibrary.wiley.com/doi/abs/10.1029/JZ069i011p02349},
	year = {1964}
}

@article{Berkner_1970,
	doi = 		{10.1088/0029-5515/10/2/006},
	url = 			{https://doi.org/10.1088/0029-5515/10/2/006},
	year = 		1970,
	month = 		{jun},
	publisher = 	{{IOP} Publishing},
	volume = 		{10},
	number = 		{2},
	pages = 		{145--149},
	author = 		{K. H. Berkner and R. V. Pyle and J. W. Stearns},
	title = 		{Cross sections for electron capture by 0.3- to 70-{keV} deuterons in {H$_2$}, {H$_2$O}, {CO}, {CH$_4$} and {C$_8$F$_{16}$} gases},
	journal = 		{Nuclear Fusion}
}

@book{SRIM_book_2015,
  	title     = 		"The Stopping and Range of Ions in Matter",
  	author    = 	{J. F. Ziegler and J. P. Biersack and M. D. Ziegler},
	publisher = 	{SRIM Co},
    address = {Chester, MD},
  	year      = 		"2015"
}

@misc{SRIM_2013,
	author = 		{James Ziegler},
	title = 		{{SRIM: The Stopping and Range of Ions in Matter}},
	url = 			{http://www.srim.org/},
	year = 		{2013}
}

@Misc{COMSOL,
 	Note  = {COMSOL Multiphysics® v. 6.2. www.comsol.com. COMSOL AB, Stockholm, Sweden.}
}

@article{Workman_2022,
	author=		{R. L. Workman and others},
	collaboration = {Particle Data Group},
	journal=	{Prog. Theor. Exp. Phys.},
  	volume=		{2022},
  	pages=		{083C01},
  	year=		{2022}
}

@article{Byrne_2022,
	author = {{Byrne, J.} and {Worcester, D. L.}},
	title = {The neutron lifetime anomaly: analysis of charge exchange and molecular reactions in a proton trap},
	DOI= "10.1140/epja/s10050-022-00786-8",
	url= "https://doi.org/10.1140/epja/s10050-022-00786-8",
	journal = {Eur. Phys. J. A},
	year = 2022,
	volume = 58,
	number = 8,
	pages = "151",
}

@article{Wietfeldt_2022,
 	title = {Comment on ``{S}earch for explanation of the neutron lifetime anomaly''},
 	author = {Wietfeldt, F. E. and Biswas, R. and Caylor, J. and Crawford, B. and Dewey, M. S. and Fomin, N. and Greene, G. L. and Haddock, C. C. and 		Hoogerheide, S. F. and 	Mumm, H. P. and Nico, J. S. and Snow, W. M. and Zuchegno, J.},
	journal = {Phys. Rev. D},
	volume = {107},
	issue = {11},
	pages = {118501},
	numpages = {4},
	year = {2023},
	month = {Jun},
	publisher = {American Physical Society},
	doi = {10.1103/PhysRevD.107.118501},
	url = {https://link.aps.org/doi/10.1103/PhysRevD.107.118501}
}

@article{Gonzalez_2021,
	title = {Improved Neutron Lifetime Measurement with $\mathrm{UCN}\ensuremath{\tau}$},
	author = {Gonzalez, F. M. and Fries, E. M. and Cude-Woods, C. and Bailey, T. and Blatnik, M. and Broussard, L. J. and Callahan, N. B. and Choi, J. H. 		and Clayton, S. M. and 	Currie, S. A. and Dawid, M. and Dees, E. B. and Filippone, B. W. and Fox, W. and Geltenbort, P. and George, E. and Hayen, L. 		and Hickerson, K. P. and Hoffbauer, M. A. and 	Hoffman, K. and Holley, A. T. and Ito, T. M. and Komives, A. and Liu, C.-Y. and Makela, M. and Morris, C. L. 	and Musedinovic, R. and O'Shaughnessy, C. and Pattie, R. W. and 		Ramsey, J. and Salvat, D. J. and Saunders, A. and Sharapov, E. I. and Slutsky, S. and 	Su, V. and Sun, X. and Swank, C. and Tang, Z. and Uhrich, W. and Vanderwerp, J. and 		Walstrom, P. and Wang, Z. and Wei, W. and Young, A. R.},
	collaboration = {$\mathrm{UCN}\ensuremath{\tau}$ Collaboration},
	journal = {Phys. Rev. Lett.},
	volume = {127},
	issue = {16},
	pages = {162501},
	numpages = {6},
	year = {2021},
	month = {Oct},
	publisher = {American Physical Society},
	doi = {10.1103/PhysRevLett.127.162501},
	url = {https://link.aps.org/doi/10.1103/PhysRevLett.127.162501}
}

@Article{Pattie_2018,
	author   = {R. W. Pattie and N. B. Callahan and C. Cude-Woods and E. R. Adamek and L. J. Broussard and S. M. Clayton and S. A. Currie and E. B. 		Dees and X. Ding and E. 	M. Engel and D. E. Fellers and W. Fox and P. Geltenbort and K. P. Hickerson and M. A. Hoffbauer and A. T. Holley and A. 			Komives and C.-Y. Liu and S. W. T. 		MacDonald and M. Makela and C. L. Morris and J. D. Ortiz and J. Ramsey and D. J. Salvat and A. Saunders and S. J. 	Seestrom and E. I. Sharapov and S. K. Sjue and Z. Tang and J. Vanderwerp and B. Vogelaar and P. L. Walstrom and Z. Wang and W. Wei and H. L. 		Weaver and J. W. Wexler and T. L. Womack and A. R. Young and B. A. Zeck},
	journal  = {Science},
	title    = {Measurement of the neutron lifetime using a magneto-gravitational trap and in situ detection},
	year     = {2018},
	number   = {6389},
	pages    = {627-632},
	volume   = {360},
 	doi      = {10.1126/science.aan8895}
}

@article{Dubbers_2021,
	author = {Dubbers, Dirk and M\"{a}rkisch, Bastian},
	title = {Precise Measurements of the Decay of Free Neutrons},
	journal = {Annual Review of Nuclear and Particle Science},
	volume = {71},
	number = {1},
	pages = {139-163},
	year = {2021},
	doi = {10.1146/annurev-nucl-102419-043156},
	URL = {https://doi.org/10.1146/annurev-nucl-102419-043156},
	eprint = {https://doi.org/10.1146/annurev-nucl-102419-043156}
}

@article{Dubbers_2011,
	title = {The neutron and its role in cosmology and particle physics},
	author = {Dubbers, Dirk and Schmidt, Michael G.},
	journal = {Rev. Mod. Phys.},
	volume = {83},
	issue = {4},
	pages = {1111--1171},
	numpages = {0},
	year = {2011},
	month = {Oct},
	publisher = {American Physical Society},
	doi = {10.1103/RevModPhys.83.1111}
}

@article{Musedinovic_2025,
  title = {Measurement of the free neutron lifetime in a magneto-gravitational trap with in situ detection},
  author = {Musedinovic, R. and Blokland, L. S. and Cude-Woods, C. B. and Singh, M. and Blatnik, M. A. and Callahan, N. and Choi, J. H. and Clayton, S. M. and Filippone, B. W. and Fox, W. R. and Fries, E. and Geltenbort, P. and Gonzalez, F. M. and Hayen, L. and Hickerson, K. P. and Holley, A. T. and Ito, T. M. and Komives, A. and Lin, S. and Liu, Chen-Yu and Makela, M. F. and O'Shaughnessy, C. M. and Pattie, R. W. and Ramsey, J. C. and Salvat, D. J. and Saunders, A. and Seestrom, S. J. and Sharapov, E. I. and Tang, Z. and Uhrich, F. W. and Vanderwerp, J. and Walstrom, P. and Wang, Z. and Young, A. R. and Morris, C. L.},
  journal = {Phys. Rev. C},
  volume = {111},
  issue = {4},
  pages = {045501},
  numpages = {11},
  year = {2025},
  month = {Apr},
  publisher = {American Physical Society},
  doi = {10.1103/PhysRevC.111.045501},
  url = {https://link.aps.org/doi/10.1103/PhysRevC.111.045501}
}

@article{Kading_2025,
  title = {Particles in finite volumes and a toy model of decaying neutrons},
  author = {K\"{a}ding, C.},
  journal = {Eur. Phys. J. C},
  volume = {85},
  pages = {758},
  year = {2025},
  month = {Jul},
  doi = {10.1140/epjc/s10052-025-14467-5}
}

@misc{data,
  title        = {Data used in writing ``{Detection of Molecular Hydrogen in a Neutron Beam Lifetime Experiment}"},
  author       = {Hoogerheide, S. F. and Biswas, R. and Caylor, J. and Crawford, B. and Dewey, M. S. and Fomin, N. and Mumm, H. P. and Nico, J. S. and Wietfeldt, F. E. and Zuchegno, J.},
  howpublished = {https://doi.org/10.18434/mds2-3859},
  year         = {2025},
}

\end{document}